\documentclass[twocolumn]{aastex631}
\usepackage{amsmath}

\newcommand{\isot}[2]{$^{#2}\mathrm{#1}$}
\newcommand{\isotm}[2]{{}^{#2}\mathrm{#1}}
\newcommand{\Xisot}[2]{$X(\isotm{#1}{#2})$}
\newcommand{\Xisotm}[2]{X(\isotm{#1}{#2})}
\newcommand{\Rconvm}{R_{\mathrm{conv}}}
\newcommand{\Rconv}{$\Rconvm$}
\newcommand{\Mconvm}{M_{\mathrm{conv}}}
\newcommand{\Mconv}{$\Mconvm$}
\newcommand{\Msun}{\ensuremath{\mathrm{M}_\odot}}
\newcommand{\Tauconvm}{\tau_{\mathrm{conv}}}
\newcommand{\Tauconv}{$\Tauconvm$}
\newcommand{\Taubetam}{\tau_{\mathrm{\beta}}}
\newcommand{\Taubeta}{$\Taubetam$}
\newcommand{\Tauecapm}{\tau_{e^{-} \mathrm{cap}}}
\newcommand{\Tauecap}{$\Tauecapm$}
\newcommand{\vrmsm}{U_{\mathrm{rms}}}
\newcommand{\vrmsrm}{\vrmsm (r)}
\newcommand{\vrms}{$\vrmsm$}
\newcommand{\vrmsr}{$\vrmsrm$}

\newcommand{\unitstyle}{\mathrm}
\newcommand{\hour}{\unitstyle{hr}} 
\newcommand{\yr}{\unitstyle{yr}}        
\newcommand{\second}{\unitstyle{s}}        
\newcommand{\erg}{\unitstyle{erg}}        
\newcommand{\gcc}{\unitstyle{g~cm^{-3}}} 
\newcommand{\kms}{\unitstyle{km~s^{-1}}} 

\newcommand{\maestro}{{\sffamily MAESTROeX}}
\newcommand{\oldmaestro}{{\sffamily MAESTRO}}

\newcommand{\microphysics}{{\sffamily Microphysics}}
\newcommand{\pynucastro}{{\sffamily pynucastro}}
\newcommand{\amrex}{{\sffamily AMReX}}
\newcommand{\yt}{{\sffamily yt}}

\begin{document}

\title{3D Convective Urca Process in a Simmering White Dwarf}

\author[0000-0002-5419-9751]{Brendan Boyd}
\affiliation{Department of Physics and Astronomy, 
Stony Brook University, Stony Brook, NY 11794-3800, USA}
\affiliation{Institute for Advanced Computational Science,
Stony Brook University, Stony Brook, NY 11794-5250, USA}

\author[0000-0001-5525-089X]{Alan Calder}
\affiliation{Department of Physics and Astronomy, 
Stony Brook University, Stony Brook, NY 11794-3800, USA}
\affiliation{Institute for Advanced Computational Science,
Stony Brook University, Stony Brook, NY 11794-5250, USA}

\author[0000-0002-9538-5948]{Dean Townsley}
\affiliation{Department of Physics and Astronomy, University of Alabama, Tuscaloosa, AL 35487-0324, USA}

\author[0000-0001-8401-030X]{Michael Zingale}
\affiliation{Department of Physics and Astronomy, 
Stony Brook University, Stony Brook, NY 11794-3800, USA}

\correspondingauthor{Brendan Boyd}
\email{boyd.brendan@stonybrook.edu}

\begin{abstract}
A proposed setting for thermonuclear (Type Ia) supernovae is a white dwarf that has gained mass from a companion to the point of carbon ignition in the core. 
In the early stages of carbon burning, called the simmering phase, energy released by the reactions in the core drive the formation and growth of a core convection zone. 
One aspect of this phase is the convective Urca process, a linking of weak nuclear reactions to convection, which may alter the composition and structure of the white dwarf. 
The convective Urca process is not well understood and requires 3D fluid simulations to properly model the turbulent convection, an inherently 3D process. 
Because the neutron excess of the fluid both sets and is set by the extent of the convection zone, the realistic steady state can only be determined in simulations with real 3D mixing processes.
Additionally, the convection is relatively slow (Mach number less than 0.005) and thus a low Mach number method is needed to model the flow over many convective turnovers. 
Using the \maestro\ low Mach number hydrodynamic software, we present the first full star 3D simulations of the A=23 convective Urca process, spanning hundreds of convective turnover times. 
Our findings on the extent of mixing across the Urca shell, the characteristic velocities of the flow, the energy loss rates due to neutrino emission, and the structure of the convective boundary can be used to inform 1D stellar models that track the longer-timescale evolution.
\end{abstract}

\section{Introduction} \label{sec:intro}

Type Ia Supernovae (SNe Ia) are extremely bright thermonuclear explosions of degenerate white dwarf material of roughly a solar mass. 
The light curves of SNe Ia are primarily powered by the decay of \isot{Ni}{56}, which enables the curves to be standardized \citep{phillips1993}. 
This feature, along with their high intrinsic brightness, makes SNe Ia ideal standard candles and a key tool in cosmology research \citep{riess1998, perlmutter1999}. 
Despite this great value, many characteristics of SNe Ia are not well understood, including the progenitor system. 
The widely studied models can generally be classified as sub-Chandrasekhar mass and Chandrasekhar mass progenitors \citep{hoeflich2006, maoz2014, liu2023}. 
Understanding the diversity of progenitor systems and their influence on the resulting SNe Ia is important for understanding the variations in SNe Ia observations.
In this paper we focus on the Chandrasekhar mass model, where a degenerate CO white dwarf accretes material from a companion star, growing to near Chandrasekhar mass before runaway carbon burning ignites near the core \citep{nomoto1984, woosley1986}. 

The earliest stage of carbon burning in the white dwarf, called the simmering phase, lasts for 1,000 to 10,000 years prior to the flame that incinerates the white dwarf \citep{woosley2004}. 
During this phase, central carbon burning drives subsonic core convection.
As the central temperature increases, the rate of carbon burning increases and works to expand the convection zone to encompass a larger portion of the white dwarf.
The rate of burning continues to increase until the burning timescale becomes shorter than the convection timescale.
At this point a flame will ignite near the center of the white dwarf and incinerate all or a portion of the white dwarf.
The exact physics of the explosion mechanism (deflagration or deflagration to detonation) continues to be an area of active research \citep{hoeflich2006, liu2023}. 

An important consequence of the simmering phase is the compositional changes due to carbon burning.
In particular, the burning generates more neutron-rich material via $\isotm{C}{12}(\isotm{C}{12}, p)\isotm{Na}{23}$ and subsequent proton and electron captures, lowering the electron fraction. 
This compositional shift can alter the nucleosynthesis of the eventual thermonuclear explosion \citep{hoyle1960, chamulak2008, piro2008}. 
1D stellar evolution models have been utilized to track total carbon burned and resulting compositional changes during the simmering phase \citep{martinez-rodriguez2016, piersanti2017, schwab2017a, piersanti2022}. 
In order to treat evolution in 1D, it is necessary to model the process of convective mixing.  
However, the appropriate way to treat convective boundaries remains an unsolved problem in stellar physics.
This problem is particularly challenging when there is significant interplay between mixing and reactive processes.
One such case in a simmering white dwarf is the convective Urca process, a linking of weak nuclear reactions to convection in the white dwarf.

Originally proposed to stabilize the carbon burning \citep{paczynski1972}, the convective Urca process leaks energy away from the star via neutrino emission from $\beta$-decay and electron capture reactions. 
Convection mixes material across a density gradient such that these reactions can occur repeatedly, and thus continuously leak energy out of the star. 
This process can lead to local cooling, and perhaps impede convection.
Further study of the convective Urca process has shown that, despite some cooling effects, it does not prevent runaway.
However, the exact impacts and importance to the simmering phase is still not well understood. 
In particular, it is unclear to what extent the convective Urca process influences the convective flow and the evolution of carbon burning during the simmering phase.

A two stream analytical approach from \cite{lesaffre2005} demonstrated that the convective Urca process may hinder/limit the strength of convection. 
Similarly, \cite{stein-wheeler2006} used a 2D implicit hydrodynamic code with boosted reaction rates to study the convective Urca process and found the convection zone was limited to the Urca shell, the region of the star bounded by the location where the $\beta$-decay and electron capture rates are equal.
However, due to the inherent differences between convection in 2D and 3D, the use of a reduced wedge geometry, as well as boosted reaction rates, the \cite{stein-wheeler2006} results should only be thought of as qualitative.
Concretely understanding the role of the convective Urca process is highly important to the use of 1D stellar evolution models, as different prescriptions of the convective Urca process can yield substantially different evolution of the simmering WD \citep{denissenkov2015, piersanti2022}.

To further investigate the convective Urca process, we present full 3D hydrodynamic simulations that can accurately model both the convective flow and the weak nuclear reactions. 
This work builds off of the analysis by \cite{willcox2018} using the \oldmaestro\ code (now \maestro\ \citealt{fan2019}).
In these simulations, we examine the extent of mixing across the Urca shell, the characteristic velocities of the flow, and the energy loss rates due to the Urca reactions.
The use of a 3D simulation is necessary to capture all the turbulent effects that are important to the convective mixing during the simmering phase. 
This turbulence is additionally vital for understanding the explosion mechanism and initial ignition of the flame in a Chandrasekhar mass explosion scenario.

In Section \ref{sec:urca}, we discuss the convective Urca process in additional detail. Section \ref{sec:sim_details} describes the \maestro\ code and the nuclear reaction network used in our simulations. 
Then, in Section \ref{sec:init_models}, we explain the initial model and conditions from which we start our simulations.
In Section \ref{sec:analysis}, we present the simulations after they have settled into a quasi-steady state. 
In Section \ref{sec:discussion}, we discuss the impacts of the velocity structure on the convective boundary and the distribution of the Urca pair in our simulations. 
And finally, in Section \ref{sec:conclusion}, we draw our conclusions and point to future work and questions.

\section{Convective Urca Process} \label{sec:urca}
The Urca process links two isotopes, called an Urca pair, via a $\beta$-decay and electron capture reaction.
The reactions work as follows for a pair of nuclei with the same atomic mass number, $A$, and proton numbers, $Z-1$ and $Z$ respectively:
\begin{equation}
    \begin{split}
        \left(Z-1,A\right)           &\longrightarrow \: \left(Z,A\right) + e^{-} + \bar{\nu}_e  \\
        \left(Z,A\right) + e^{-} &\longrightarrow \: \left(Z-1,A\right) + \nu_e
    \end{split}
\end{equation}
A direct result of the Urca process is some local cooling due to neutrino emission in each weak reaction.
In a degenerate white dwarf, these weak reactions depend on temperature and the electron density. 
In most regions only one reaction will be active, i.e.\ electron captures at higher densities and $\beta$-decays at lower densities. 
The transition between these regions, where the reaction rates are equal, is called the Urca shell.
The key to the convective Urca process is that convection transports underlying material outward past the Urca shell, and vice versa. 
As convection cyclically mixes material back and forth across the shell, the Urca process will occur continuously, so a single nucleus can undergo repeated reactions. 
The cyclical nature enables the convective Urca process to have a meaningful impact even with relatively small abundances of the Urca pair. 

There are many Urca pairs relevant to simmering white dwarfs, but we focus this paper on the $A=23$ Urca pair, \isot{Na}{23} -- \isot{Ne}{23}. 
The pair is linked by the weak reactions:

\begin{equation} \label{eqn:urca}
    \begin{split}
        \isotm{Ne}{23}           &\rightarrow \: \isotm{Na}{23} + {e}^{-} + {\Bar{\nu}}_{e} \\
        \isotm{Na}{23} + {e}^{-} &\rightarrow \: \isotm{Ne}{23} + {\nu}_{e}
    \end{split}
\end{equation}
\isot{Na}{23} is relatively abundant in the white dwarf at the onset of carbon burning, $\Xisotm{Na}{23} \approx 10^{-4}$ \citep{martinez-rodriguez2016, piersanti2017, schwab2017a}. 
And the $A=23$ Urca shell is approximately located at a density $\rho_{\rm{urca}} \sim 1.7 \times 10^9 \, \gcc$ \citep{suzuki2016}. 
Further references to an Urca pair in this paper are in relation to the $A=23$ pair unless otherwise specified. 
We note that by including only one Urca pair, our simulations may underrepresent the full impact of the convective Urca process. 
Recent work \citep{piersanti2022} has demonstrated the inclusion of additional Urca pairs, such as \isot{Ne}{21} -- \isot{F}{21}, may be vital to understanding the simmering phase.

While it is clear how vigorous convection continually drives the Urca process in a simmering white dwarf, the more subtle feature of the convective Urca process is how the Urca process affects convection.
The Urca reactions alter the electron fraction which impacts the density/pressure in the degenerate white dwarf.
In particular, the electron capture material (largely located interior to the Urca shell) is denser than the $\beta$-decay material (largely located exterior to the Urca shell) at the same pressure.
This discrepancy can produce a dragging effect that may slow down convection.
Additionally, the convective Urca process can produce compositional gradients near the Urca shell which work to hinder convection.
Studies of 2D simulations \citep{stein-wheeler2006} and analytic approaches \citep{lesaffre2005} have demonstrated that the convective Urca process may slow the convective flow and effectively reduce the total kinetic energy in the white dwarf. 
To test and properly characterize these theories and effects, a 3D hydrodynamic model is needed to simulate the turbulent convection linked with the Urca reactions.

\section{Numerical Methodology} \label{sec:sim_details}
We present simulations generated using \maestro\ \citep{fan2019}, a massively parallel low Mach number hydrodynamic code built on the \amrex\ framework \citep{zhang2019}. 
\maestro\ is specifically designed to study stellar interiors and atmospheres.
Many stellar processes, such as core convection, occur in low Mach number environments (i.e.\ the sound speed is fast compared to the characteristic fluid velocity).
When modeling low Mach number flow, a standard compressible finite-volume code is limited by the acoustic wave timescale as opposed to the characteristic timescale of the flow itself.
This limit on the timestep makes modeling slow moving fluids inefficient and often impossible for very low Mach numbers.
Our presented simulations are limited by a Mach number of ${\sim} 0.005$. 

\maestro\ solves the following low Mach number hydrodynamic equations which correspond to the evolution of mass, momentum and energy:

\begin{equation} \label{eqn:mass}
    \frac{\partial(\rho X_k)}{\partial t} = - \nabla \cdot (\rho X_k \mathbf{U}) + \rho \dot{\omega}_k
\end{equation}

\begin{equation} \label{eqn:momentum}
    \frac{\partial \mathbf{U}}{\partial t} = - \mathbf{U} \cdot \nabla \mathbf{U} - \frac{\beta_0}{\rho} \nabla \left( \frac{p - p_0}{\beta_0} \right) - \frac{\rho - \rho_0}{\rho} g \mathbf{e}_r
\end{equation}

\begin{equation} \label{eqn:energy}
    \frac{\partial (\rho h)}{\partial t} = - \nabla \cdot ( \rho h \mathbf{U} ) + \rho H_{\mathrm{nuc}}
\end{equation}

Where $\rho, \mathbf{U}, h$ are mass density, velocity and enthalpy respectively. 
$X_k$ is the mass fraction of the $k$th isotope and $\dot{\omega}_k$ is the creation/destruction rate of that isotope. 
Note that $X_k$'s are defined such that $\sum_k X_k = 1$.
$\rho_0$, $p_0$ correspond to the base state, essentially the angle-averaged background state of the star, which remains in hydrostatic equilibrium. 
$\beta_0$ is a density-like variable that captures the background stratification.
And finally, $H_\mathrm{nuc}$  is the specific energy generation rate due to nuclear reactions.
This formulation enforces conservation of total energy at low Mach number in the absence of external heating or viscous terms  \citep{klein2012, vasil2013}.
Note that in the presented simulations we do not evolve the base state, which eliminates some time-dependent terms from the above equations, as well as Eqn.\ \ref{eqn:div} (see \citealt{fan2019} for full equation set).

The equation of state (EOS) is recast to a divergence constraint (analogous to incompressible and anelastic approximations).
After the velocity is evolved using Eqn.\ \ref{eqn:momentum}, it is projected to satisfy the constraint: 

\begin{equation} \label{eqn:div}
    \nabla \cdot (\beta_0 \mathbf{U}) = \beta_0 S 
\end{equation}

Here, $S$ is a source term that accounts for perturbations related to compositional changes and heating from reactions, while $\beta_0$ accounts for the effects of stratification.
Through this projection method, acoustic waves are effectively filtered out, allowing \maestro\ to track the advection timescale and accurately model slow moving flows like core convection. 
Overall, the algorithm is second-order accurate in space and time.

\maestro\ solves these hydrodynamic equations on a structured mesh handled by \amrex. 
In the presented simulations we use a ``full-star" geometry, which places the star at the center of a large 3D cartesian grid.
Through \amrex, block-structured adaptive mesh refinement is implemented to allow for varying levels of refinement on the grid.
This allows for a higher ``effective resolution", where unimportant areas of the grid use lower resolution, minimizing computational costs.
Although the refinement can be implemented in an adaptive way (i.e.\ adjusting to changes in the simulation), in the presented simulations we use a static approach in which the refinement layers do not change from the initial settings (see Section \ref{sec:init_models} for detailed description). 
For further details on \maestro\ and the low Mach number algorithm see \cite{fan2019} and references within.

\begin{figure}
    \centering
    \plotone{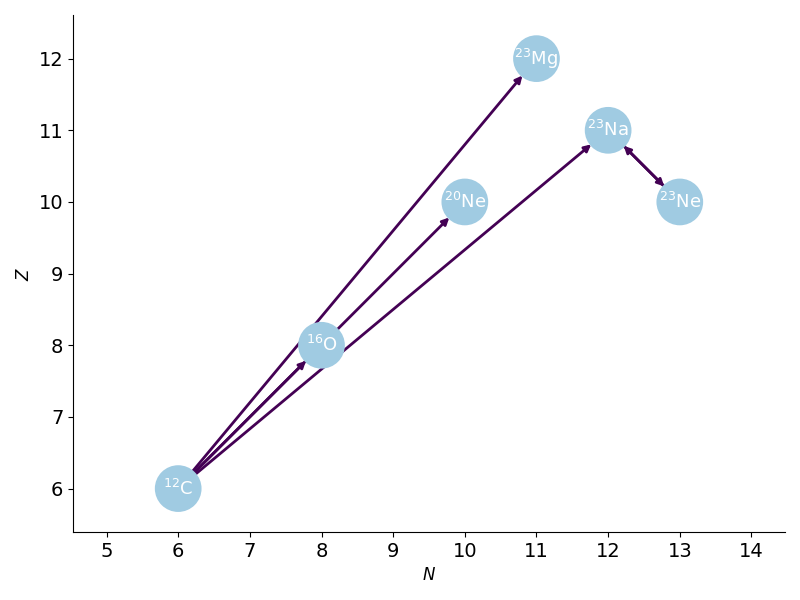}
    \caption{\label{fig:urca_net} 
        A graphical visualization of the nuclear reaction network used in the simulations. 
        Each node represents a different isotope in the network.
        Arrows represent reactions connecting two isotopes. 
        The direction of the arrow indicates the forward direction of the reaction.
        The horizontal axis indicates the neutron number. 
        The vertical axis indicates the proton number. 
        Helium, protons, and neutrons are in the network, but are excluded from this plot for the sake of clarity.
    }
\end{figure}

A critical piece to modeling the convective Urca process is the coupling of nuclear reactions to the fluid motions.
In \maestro\ this is typically done via Strang-splitting that couples the reaction rate equations to the fluid equations.
The nuclear network used is shown in Figure \ref{fig:urca_net} and was generated using the python library \pynucastro\ \citep{smith2023}.
The network incorporates a simple carbon burning network with the addition of the A=23 Urca reactions (see Eqn. \ref{eqn:urca}).
The included rates associated with carbon burning come from the JINA REACLIB database \citep{cyburt2010} and are listed below:
\begin{equation}
     \begin{split}
        \isotm{C}{12}  + \isotm{C}{12} &\rightarrow \: \isotm{Ne}{20} + \isotm{He}{4}   \\
        \isotm{C}{12}  + \isotm{C}{12} &\rightarrow \: \isotm{Mg}{23} + n               \\
        \isotm{C}{12}  + \isotm{C}{12} &\rightarrow \: \isotm{Na}{23} + p               \\
        \isotm{C}{12}  + \isotm{He}{4} &\rightarrow \: \isotm{O}{16}                    \\
        n                              &\rightarrow \: p + {e}^{-} 
    \end{split}   
\end{equation}
We incorporate the effects of Coulomb screening on the carbon burning rates following \cite{graboske1973} for the weak limit and \cite{alastuey1978, itoh1979} for the strong limit.
For the Urca reactions we use tabulated rates from \cite{suzuki2016} with bilinear interpolation.
And we account for the thermal neutrino losses from the hot plasma following \cite{itoh1996}.
We use a publicly available general purpose stellar EOS described in \cite{timmes2000}, which takes into account the contributions of ions, electrons and radiation. 
Implementations of the reaction network, Coulomb screening effects, thermal neutrino losses, and the EOS in C++ are developed as a part of the \microphysics\ project \citep{microphysics2024} which supports \amrex\ based simulation codes.

In the low density outer regions of the simulation, we use a velocity sponge to dampen gravity wave excitations caused by the core convection, as described in \citep{zingale2009}.
The sponge dampens the velocities toward zero by dividing by a constant factor, in our case $\kappa = 10$, for densities less than $\rho_{\mathrm{sponge}} = 10^6 \, \gcc$.
The sponge ensures that these low density regions, far outside the convection zone, do not limit our timestep.
The addition of this velocity sponge does not impact the convection zone, where densities are of order $10^9 \, \gcc$, and thus does not impact the conclusions we make from these simulations.

\begin{figure}
    \centering
    \plotone{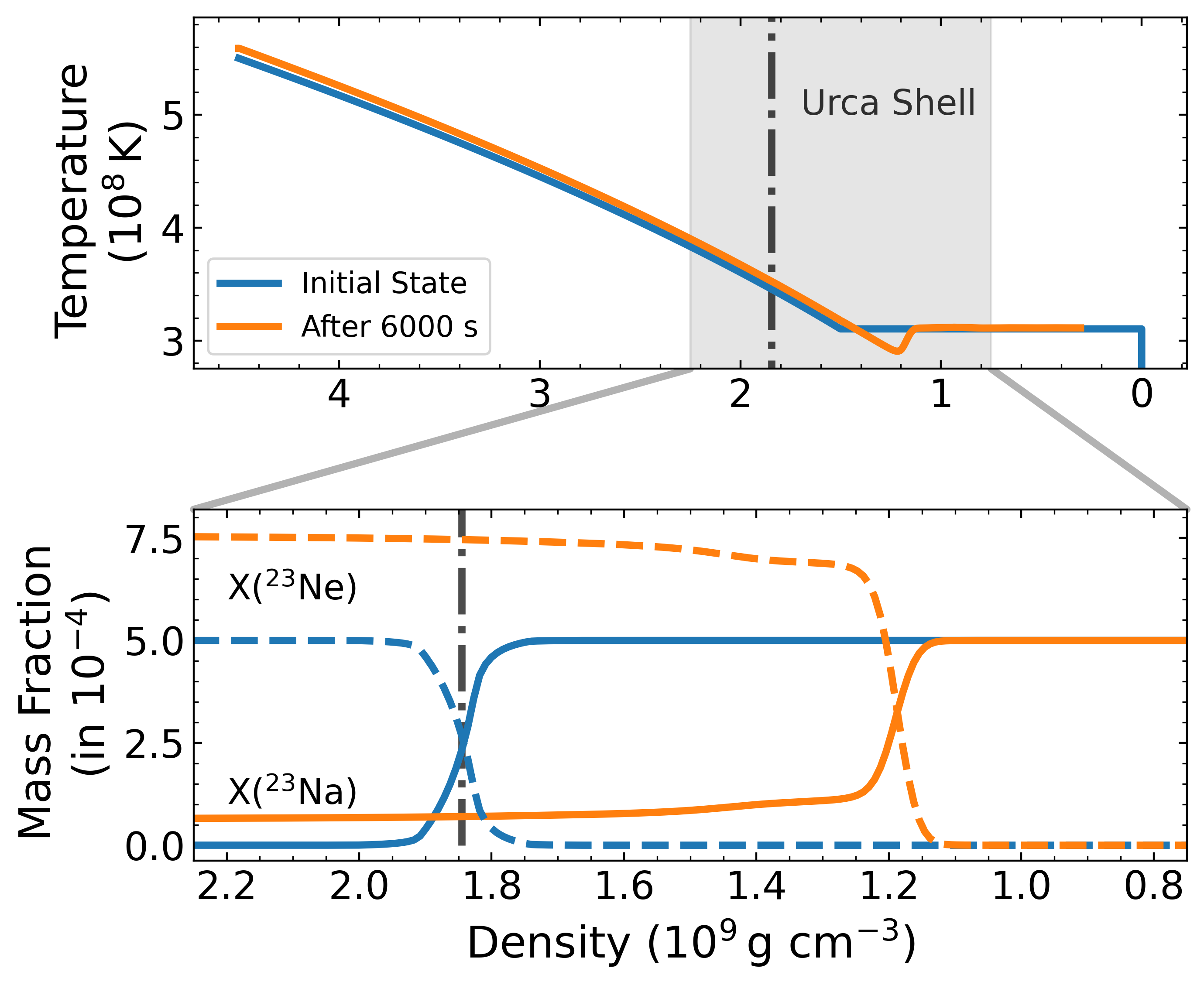}
    \caption{\label{fig:init_prof} 
        The top plot shows the temperature vs density profile of the star at the initial time (blue curve) and at the end of the high resolution simulation (orange curve). 
        The black vertical line indicates the location of the Urca shell. 
        The bottom plot shows the distribution of \isot{Ne}{23} (dashed curves) and \isot{Na}{23} (solid curves) for the shaded region of the star. 
        The blue curves represent the initial state and the orange curves represent the end of the high resolution simulation. 
    }
\end{figure}

\section{Initial Models} \label{sec:init_models}

We present two simulations with effective resolutions of 5.0 km and 2.5 km. 
From here on, we refer to these as the low resolution simulation and high resolution simulation respectively.
Besides the change in resolution, the two simulations use the same numerical methodology described above and use the same initial model described below (except where specified).

We construct a simple, parameterized model of a simmering white dwarf with an isentropic core and isothermal envelope (see blue curves in Figure \ref{fig:init_prof}). 
This initial model, as well as a variety of other models for \amrex\ based astronomical codes, is developed and maintained by \cite{initial_models2024}. 
This initial model was originally investigated by \cite{willcox2018} and was motivated by the 1D simmering white dwarf simulations in \cite{martinez-rodriguez2016}.
We construct the model by first setting the central density, $\rho_{\rm{c}} = 4.5 \times 10^9 \, \gcc$, and temperature $T_{\rm{c}} = 5.5 \times 10^8 \ \mathrm{K}$, then integrating outward while maintaining hydrostatic equilibrium.
We specify the temperature profile to follow an adiabatic curve for the first $0.5 M_{\odot}$ in mass coordinates, which sets the approximate initial size of the convection zone.
For the remainder of the star, we then specify an isothermal profile, which is convectively stable.
Since there is no a priori method for knowing the extent of the convection zone, the $\Mconvm = 0.5 \, \Msun$ was chosen so that the zone extended just past the Urca shell.
The Urca shell is located at approximately $R_{\mathrm{urca}} = 415 \, \mathrm{km}$, and the initial isentropic core extends an additional $50 \, \mathrm{km}$ radially outward.

This choice in initial \Mconv\ is decidedly smaller than the roughly $1 \, \Msun$ convection zone predicted by \cite{martinez-rodriguez2016}.
We chose a smaller convection size, which modestly surpasses the Urca shell, with the understanding that the convective Urca process may limit the size of the convection zone.
Comparing our initial model to more recent work, the LED model discussed in section 5.2 of \cite{piersanti2022} found the A=23 Urca pair constrained the convection zone to the Urca shell.
The evolution of the LED model, compared to \cite{martinez-rodriguez2016}, differs in part because of the constrained convection zone.
Thus an initial model based on the LED model will differ structurally from the one presented in Figure \ref{fig:init_prof}.
Note that \cite{piersanti2022} extends their analysis to more Urca pairs, specifically the significant role \isot{Ne}{21} -- \isot{F}{21} may have on the simmering phase.
The inclusion of more Urca pairs in future simulations is discussed further in Section \ref{sec:conclusion}.
We discuss this choice of initial \Mconv\ and the consequences on the simulation in further detail in Section \ref{sec:discussion}.

Additionally, we seed a velocity perturbation near the central region by summing 27 Fourier modes with a maximum amplitude of ${\sim} 4 \times 10^{-3} \, \kms$ for the low resolution simulation and ${\sim} 4 \, \kms$ for the high resolution simulation (see \citealt{zingale2009} for greater detail).
These perturbations are contained to the inner $250 \, \mathrm{km}$ in radius.
This initial perturbation has three purposes. 
Firstly, the initial velocity helps move energy away from the center, preventing a carbon burning runaway effect due to the initial static nature of the simulation.
Secondly, initializing a non-zero velocity helps with the convergence of the initial projection onto the constraint equation (Eqn.\ \ref{eqn:div}) in the \maestro\ algorithm.
And lastly, the perturbation helps jump start the convection so that we can more quickly reach a quasi-steady state. 

This jump start to developing convection is why we increased the amplitude of the perturbation for the high resolution simulation. 
The initial velocity seed is quickly erased after just a few convective turnovers, even with an increased amplitude.
Once the convection zone is established, the average speed is up to an order of magnitude greater than the initial amplitude of the Fourier seeds, see Section \ref{subsec:conv_zone} for more details on the characteristic speed in the convection zone. 
Additionally, the size of the convection zone quickly extends well past the region where the velocity perturbation is defined.

The initial composition is largely uniform with $\Xisotm{C}{12} = 0.39975$ and $ \Xisotm{O}{16}= 0.59975$ with a trace amount of the A=23 Urca pair, $\Xisotm{Na}{23} + \Xisotm{Ne}{23} = 5 \times 10^{-4}$, that is constant throughout the white dwarf. 
We distributed the Urca pair such that the Urca reactions were in local equilibrium (see blue curves in bottom plot of Figure \ref{fig:init_prof}).
Note that this is not the true dynamic equilibrium, but instead the distribution of the Urca pair if there were not convective mixing across the Urca shell.

\maestro\ maps these 1D profiles (density, temperature, pressure etc.) onto a non-uniform 3D cartesian grid via interpolation, see \cite{fan2019} for greater details. 
For the high (low) resolution simulation, the cartesian grid is a $5120 \, \unitstyle{km}$ wide cube split into four (three) refinement layers. 
The coarsest layer has a resolution of $20 \, \unitstyle{km}$ and the finest layer resolves to $2.5 \, \unitstyle{km}$ ($5.0 \, \unitstyle{km}$). 
We ensure that the finest layer of resolution fully encompasses the isentropic core, which prevents any changes in resolution from potentially impacting the turbulent convective motions.

\section{Analysis} \label{sec:analysis}
In this section, we split out our various analyses to explore three general properties of the simulations. 
In subsection \ref{subsec:conv_zone}, we analyze the broad properties of the convection zone, including the size and growth of the convection zone, estimating the convective turnover time, and identifying an energy cascade due to turbulent fluid motions. 
We then analyze the distribution of turbulent motions throughout the convecting region and the impact of the growing convection zone on relevant temperature gradients.
In subsection \ref{subsec:vel_struct}, we identify the 3D velocity structure in the convection zone, the ways this structure evolves with time, and investigate the subsequent impact on the convective mixing.
Lastly, in subsection \ref{subsec:urca_proc}, we analyze the convective Urca process including the characteristic timescales and the energy losses related to the neutrino cooling.

\subsection{Convection Zone Properties} \label{subsec:conv_zone}

\begin{figure}
    \centering
    \plotone{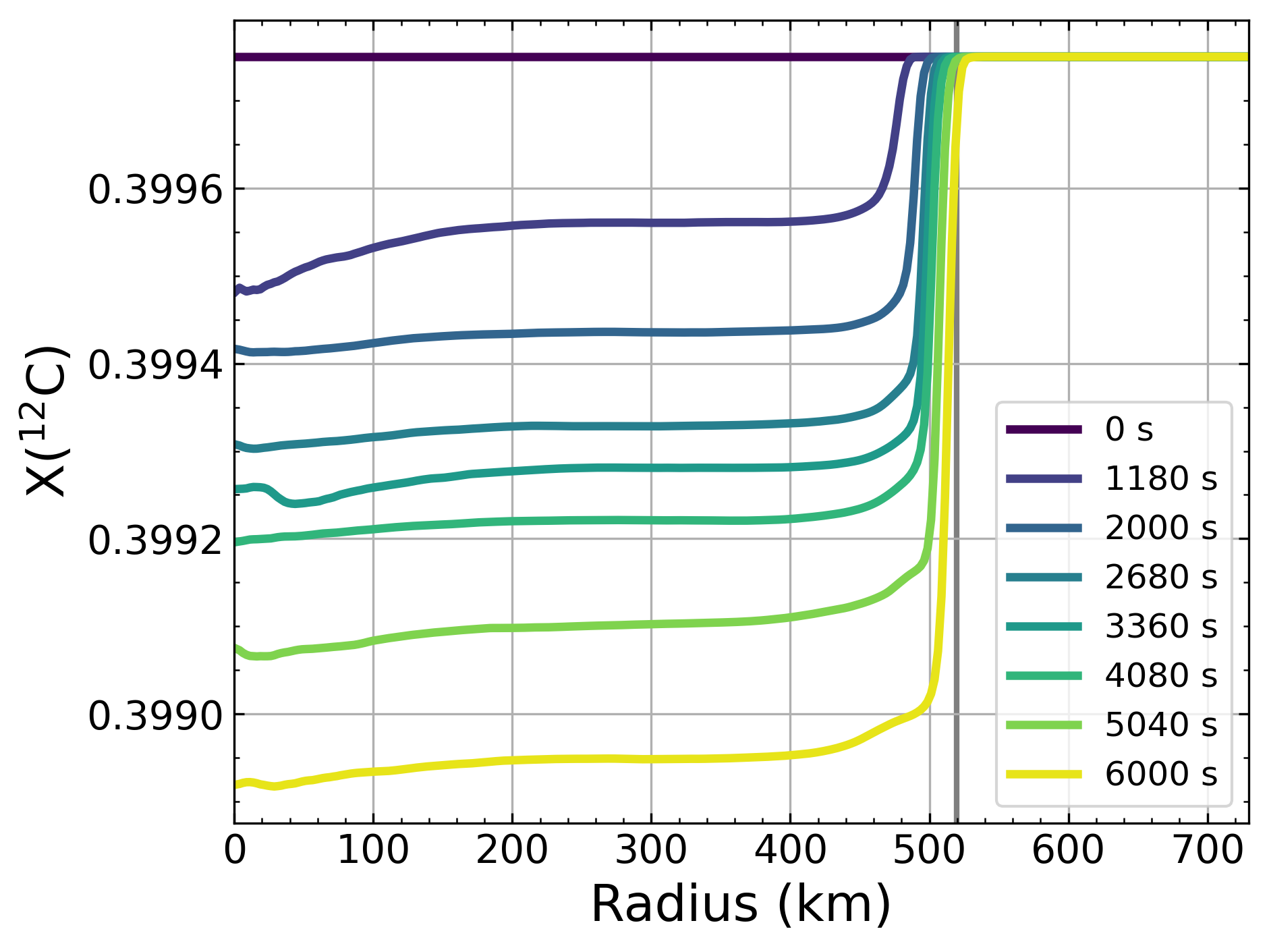}
    \caption{\label{fig:c12_profiles}
        The angle-averaged \Xisot{C}{12} vs bin radius for the high resolution simulation. 
        Each curve represents a distinct snapshot in time indicated by the color. 
        Darker blue represents early times and lighter greens represent later times.
        The grey vertical line at about $520 \, \mathrm{km}$ marks \Rconv\ at the end of the simulation.
        }
\end{figure}

The low resolution and high resolution simulations were run to about $9000 \, \second$ and $6000 \, \second$ respectively, which corresponds to an order of a hundred convective turnovers.
During the simulation, carbon burning in the center of the star drives convection, which efficiently mixes the products of this burning throughout the convection zone.
This leads to a uniformly lower carbon fraction in the convection zone as compared to the exterior layers, see the radial profiles of \Xisot{C}{12} in Figure \ref{fig:c12_profiles}.
These profiles of \Xisot{C}{12} are mass weighted angle-averages for a given radial bin.

We define the radial size of the convection zone, \Rconv, based on the peak gradient of these \Xisot{C}{12} radial profiles (following a similar approach to \citealt{andrassy2022}). 
To calculate this radius, we first construct a discrete gradient from these \Xisot{C}{12} profiles using a central difference formulation.
From this discrete gradient, we find the maximum value and fit a parabola to the nearest 7 points (3 on each side).
Finally, we define \Rconv\ from the peak of this fitted parabola.
By fitting a parabola to the nearby points, we smooth out some of the noise due to our radial discretization. 

\begin{figure}
    \centering
    \plotone{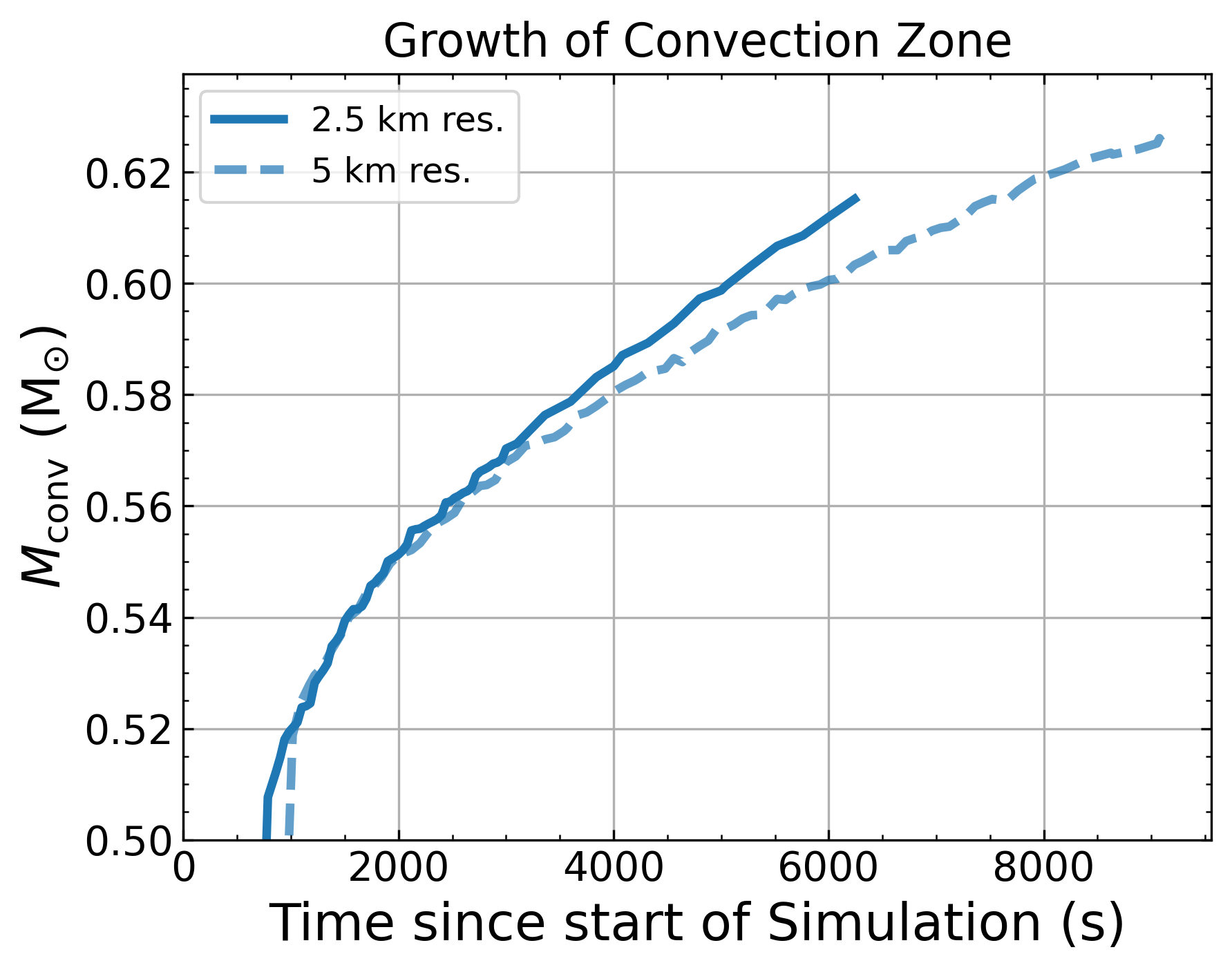}
    \caption{ \label{fig:conv_mass}
        Time variation of the total mass of the convection zone.
        The solid curve represents the high resolution simulation (2.5 km) while the dashed curve represents the low resolution simulation (5 km).
        }
\end{figure}

In our simulations, we observe the convection zone continues to expand, even after the simulations settle from an initial transitory period (the first ${\sim} 3000 \, \second$) caused by our nearly static initial conditions.
This growth is directly seen in the shifting \Xisot{C}{12} profiles in Figure \ref{fig:c12_profiles}. 
To quantify this growth, we calculate the mass contained in the convection zone, \Mconv, throughout the simulation's runtime (see Figure \ref{fig:conv_mass}).
Some expansion of the convection zone is expected due to the energy deposited by carbon burning. 
In addition, convective overshooting and instabilities at the interface of the convection zone and stable envelope, collectively labeled convective boundary mixing \citep{meakin2007, denissenkov2013, gilet2013, herwig2023}, work to further expand the convection zone.
However, a linear extrapolation of the growing convection zone, after the initial transitory period, indicates a $\dot{M}_{\mathrm{conv}} \approx 0.03 {-} 0.05 \; \Msun~\hour^{-1}$. 
This growth is alarmingly quick.
The expected growth should be a rate of only hundreths of $\Msun~\yr^{-1}$ or less, with the convection zone slowly encompassing most of the interior of the star just prior to flame ignition.

This extreme growth indicates the convection zone is not yet at a stable size. 
The parameterized initial model set a relatively small convection zone ($\Mconvm \sim 0.5 \, \Msun$) with a relatively hot isothermal envelope ($3 \times 10^8 \,  \mathrm{K}$).
For the given central density and temperature, a convection zone ${\sim} 1 \, \Msun$ and isothermal region of ${\sim}1 \times 10^8 \, \mathrm{K}$ would be more in line with the fiducial model presented in Figure 1 of \cite{martinez-rodriguez2016}. 
A major caveat in comparing to stellar models, like that presented in \cite{martinez-rodriguez2016}, is the limitations of stellar models to accurately describe convective mixing.
These prescriptions of convective mixing do not incorporate the effects of the Urca process and cannot describe the temperature valley seen in Figure \ref{fig:init_prof}.
We leave further discussion of the initial model and the relation between convection and the temperature valley to Section \ref{sec:discussion}.
Ultimately, further analysis and testing is needed to construct a more realistic initial model that will yield a stable convection zone under these conditions.

\begin{figure}
    \centering
    \plotone{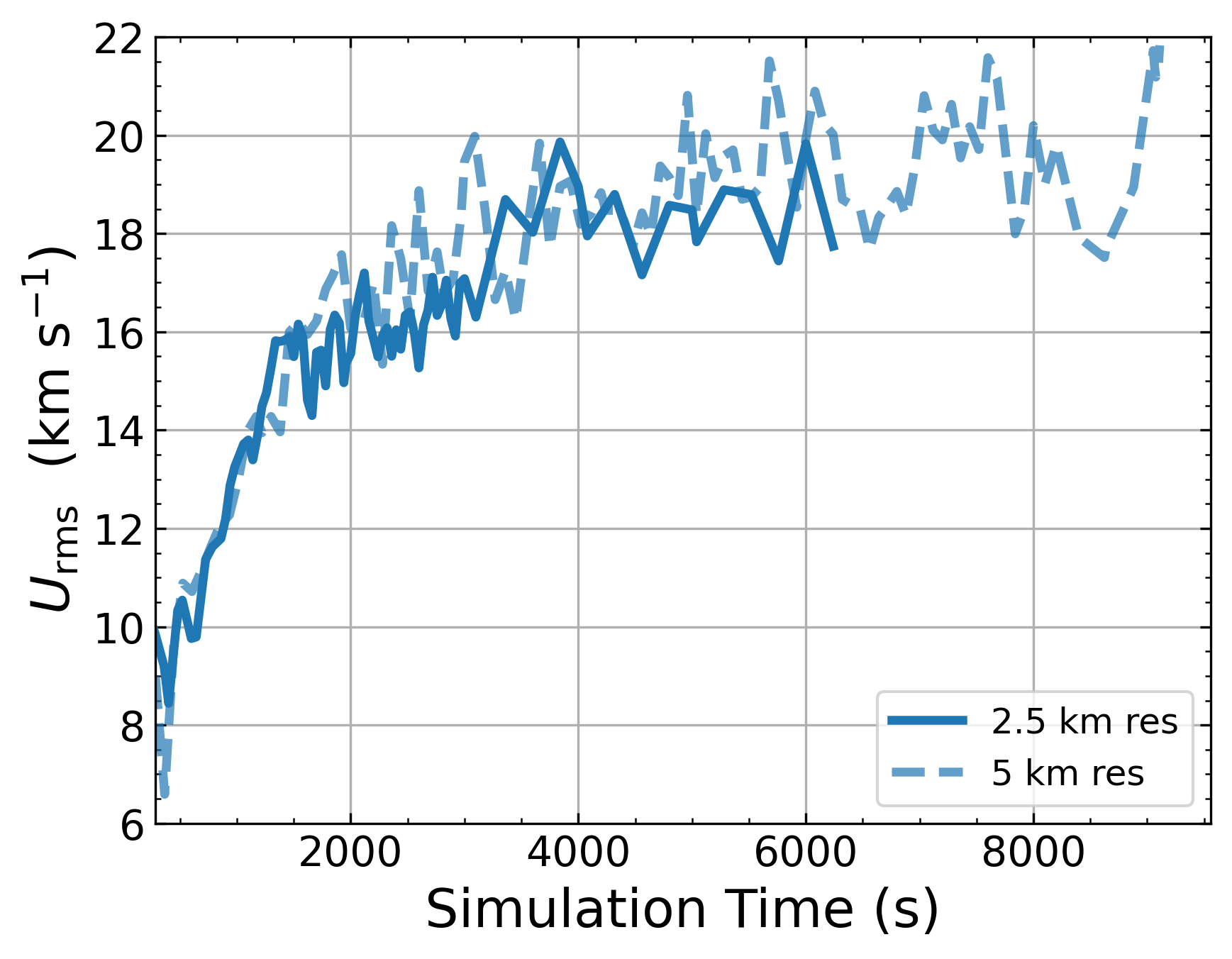}
    \caption{\label{fig:vrms_conv}
        Time variation of the rms velocity in the convection zone.
        The solid curve represents the high resolution simulation (2.5 km) while the dashed curve represents the low resolution simulation (5 km).
    }
\end{figure}

We determine the characteristic convective velocity by calculating the rms velocity, \vrms, in the convection zone as:

\begin{equation} \label{eqn:vrms}
    \vrmsm = \sqrt{ \langle {U_x} ^2 \rangle + \langle {U_y} ^2 \rangle + \langle {U_z} ^2 \rangle }
\end{equation}
where $U_x, U_y, U_z$ are velocity components in the $x,y,z$ direction. 
The angled brackets $\langle \rangle$ represent a density weighted average over the convection zone.
For example, the x component is calculated:
\begin{equation}
    \langle {U_{x}}^2  \rangle= \left( \sum_{\mathrm{conv}}  \rho {U_x}^2 \right) / \sum_{\mathrm{conv}} \rho
\end{equation}
where we loop over all cells in the convection zone, defined by \Rconv.
We track the evolution of \vrms\ in the convection zone vs time in Figure \ref{fig:vrms_conv}.
Note that we only start plotting after about $100 \, \mathrm{s}$ of simulation time, as \Rconv\ is ill-defined prior to this.
Comparing the two resolutions, we can see the increased amplitude of the initial velocity perturbation (as discussed in Section \ref{sec:init_models}) has a relatively minor impact on the evolution of \vrms.
After the initial $3000 \, \second$ transitory period, \vrms\ largely steadied out to around $19 \, \kms$, with fluctuations of a few $\kms$.
This suggests a largely established, albeit still growing, convection zone for the remainder of the simulation. 

Using \Rconv\ and \vrms\ as characteristic length scales and speed scales, we estimate the characteristic turnover timescale of the convection zone, \Tauconv, as:

\begin{equation}
    \Tauconvm = 2 \frac{\Rconvm}{\vrmsm}
\end{equation}
This estimate yields a turnover timescale of $\Tauconvm \sim 50 \, \second$ for our simulation.

\begin{figure}
    \centering
    \plotone{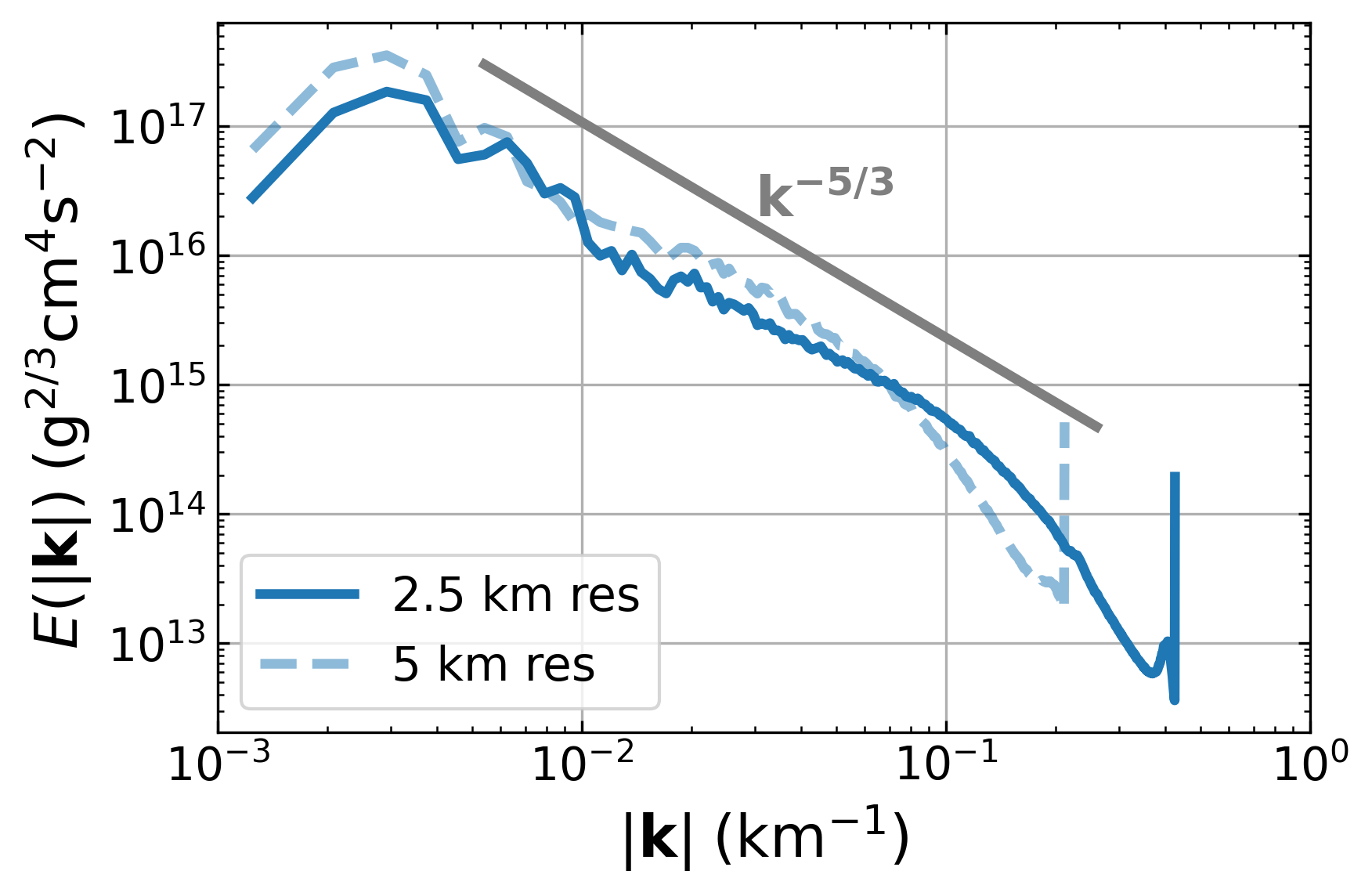}
    \caption{ \label{fig:powerspec}
        The generalized kinetic energy power spectrum for the high resolution (solid blue) and the low resolution (dashed blue) simulations. 
        A -5/3 scaling curve is plotted in grey for reference.
    }
\end{figure}

To verify the flow in the convection zone is turbulent, we calculate the kinetic energy power spectrum after the simulation has reached the quasi-steady state.
This spectrum calculation is limited to a cube with side-length of $1200 \, \mathrm{km}$ which fully encompasses the convection zone.
Due to density stratification, it is important to weight the velocity appropriately by $\rho^{1/3}$ to get a Kolmogorov-type five-thirds decay law (as shown in \citealt{kritsuk2007, nonaka2012}). 
With this weighting, we calculate a generalized energy spectrum:

\begin{equation}
    E(|\mathbf{k}|) =  \int_{\mathbf{S}(|\mathbf{k}|)} \frac{1}{2} \mathbf{\hat{V}}(\mathbf{k}) \mathbf{\hat{V}^{*}}(\mathbf{k}) \, d\mathbf{S} 
\end{equation}
where $\mathbf{V} = \rho^{1/3} \textbf{U}$ and $\mathbf{\hat{V}}$ is the associated Fourier transform, with ${}^*$ denoting the complex conjugate.
$\mathbf{S}(|\mathbf{k}|)$ is the surface of constant $| \mathbf{k} |$.
The resulting power spectrum, see Figure \ref{fig:powerspec}, indicates Kolmogorov-type scaling over multiple orders of magnitude. 
This indicates turbulent behavior is transporting energy from large length scales down to smaller length scales.
The higher resolution curve (solid blue) indicates five-thirds scaling down to smaller length scales, which indicates the benefit of increasing resolution for this problem.

\begin{figure}
    \centering
    \plotone{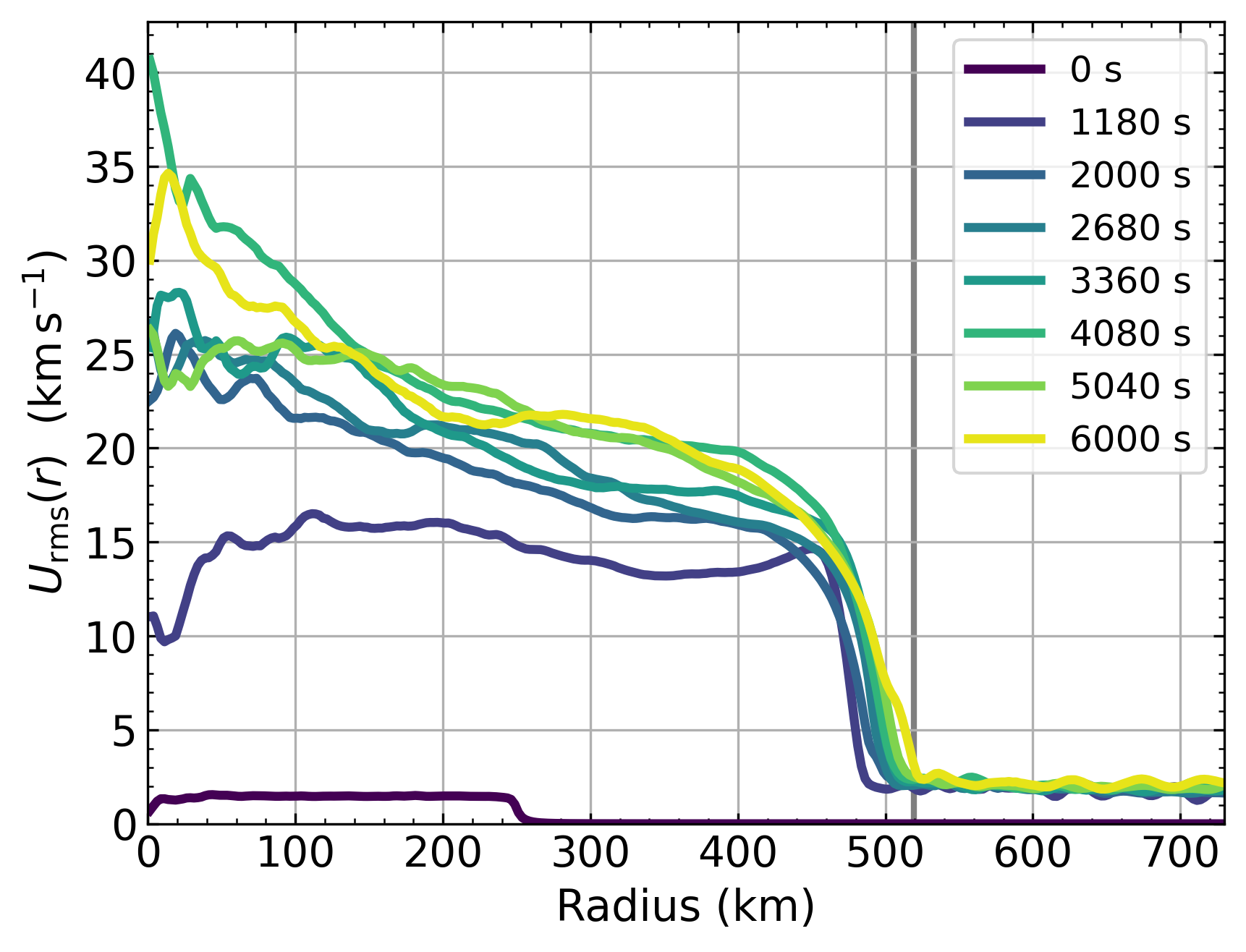}
    \caption{\label{fig:vrms_profiles}
        \vrmsr\ vs bin radius for the high resolution simulation. 
        Each curve represents a distinct snapshot in time indicated by the color. 
        Darker blue represents early times and lighter greens represent later times.
        The grey vertical line at about $520 \, \mathrm{km}$ marks \Rconv\ at the end of the simulation.
        }
\end{figure}

\begin{figure}
    \centering
    \plotone{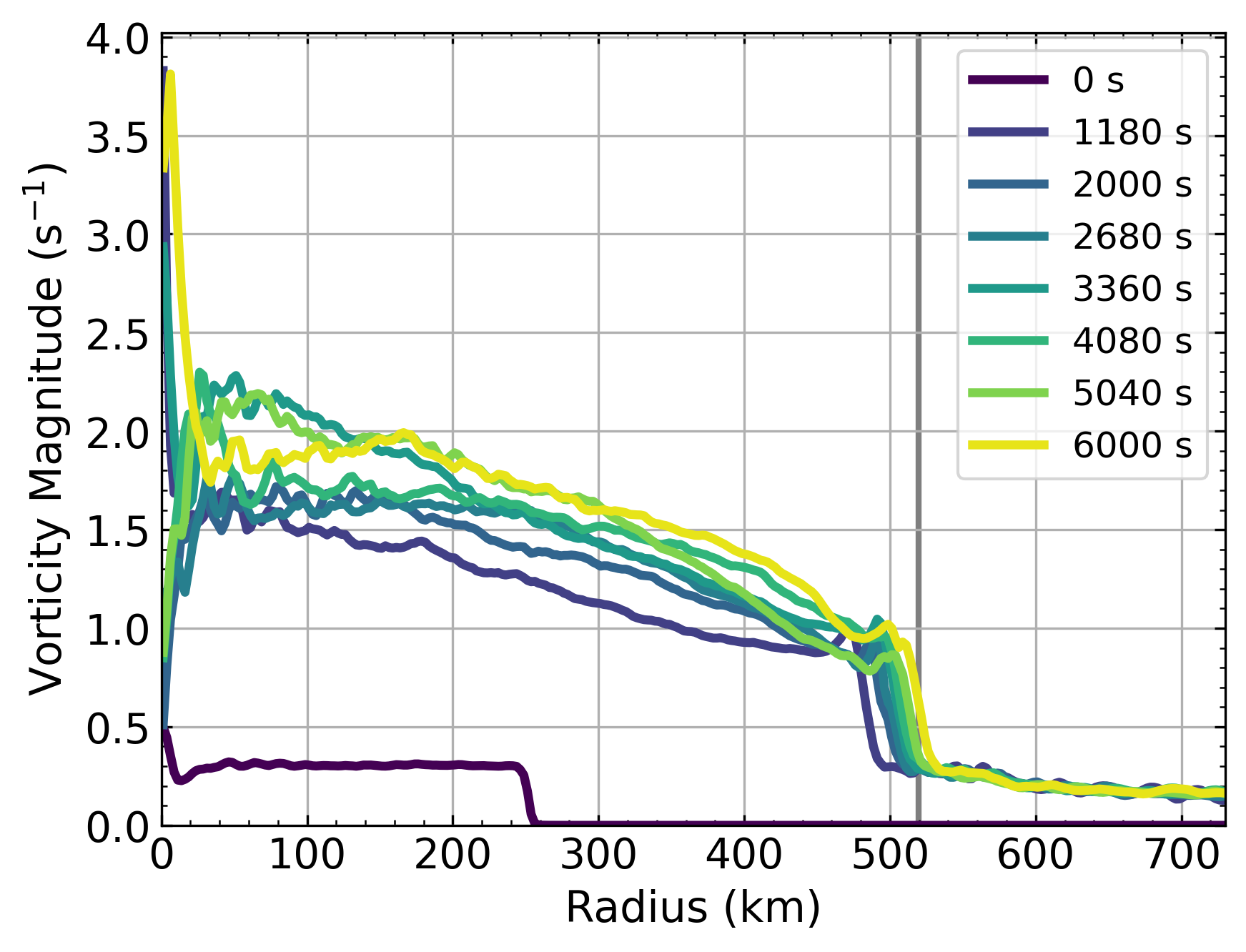}
    \caption{\label{fig:vort_profiles}
        The angle-averaged vorticity vs bin radius for the high resolution simulation. 
        Each curve represents a distinct snapshot in time indicated by the color. 
        Darker blue represents early times and lighter greens represent later times.
        The grey vertical line at about $520 \, \mathrm{km}$ marks \Rconv\ at the end of the simulation.
        }
\end{figure}

To further investigate the extent of turbulent motions and mixing in different regions of the convection zone, we calculate the radial profiles of the rms velocity and the average vorticity magnitude, similar to Figure \ref{fig:c12_profiles}.
We calculate the rms velocity in radial shells, \vrmsr, analogously to Eqn. \ref{eqn:vrms}, with the exception that the summation is over a radial bin centered at radius $r$, as opposed to the whole convection zone. 
In Figure \ref{fig:vrms_profiles}, we show the \vrmsr\ profile at a series of times throughout the simulation. 
The distribution of the rms velocity is more complex and varied when compared to the relatively smooth \Xisot{C}{12} profiles in Figure \ref{fig:c12_profiles}.
For $r < 400 \, \mathrm{km}$, we find that \vrmsr\ initially increases in the convection zone during the first $3000 \, \second$ transitory period, before fluctuating around ${\sim}25 \, \kms$ for the remainder of the simulation.
Generally, \vrmsr\ has an initial downward slope that then steepens near $500 \, \mathrm{km}$.
At the convective boundary, \vrmsr\ has already decreased closer to ${\sim}15 \, \kms$ from its maximum value near the center of the star.
Outside the convection zone, $r > 520 \, \mathrm{km}$, \vrmsr\ drops to about $2 \, \kms$.

To complement the \vrmsr\ profiles, we also plot the vorticity magnitude vs radius in Figure \ref{fig:vort_profiles}.
These profiles are density weighted angle-averages of the vorticity magnitude.
Similar to the \vrmsr, we find that the vorticity magnitude increases in the convection zone before settling to a relatively uniform shape (though there are some large fluctuations at small radii).
Like the \vrmsr\ profiles, the magnitude of the vorticity consistently decreases from the center out to around $500 \, \mathrm{km}$, before a bump in the magnitude precedes a steep drop off to about $0.2 \, \mathrm{s}^{-1}$.
The bump, and steep drop, occurs at a radius consistent with our definition of \Rconv\ from the \isot{C}{12} profile.
This is most clearly seen for the $t = 6000 \, \mathrm{s}$ profile where the bump occurs just a few km before \Rconv, denoted by the grey vertical line.
Further analysis of the 3D structure of the turbulent velocity field is done in Section \ref{subsec:vel_struct}.

\begin{figure}
    \centering
    \plotone{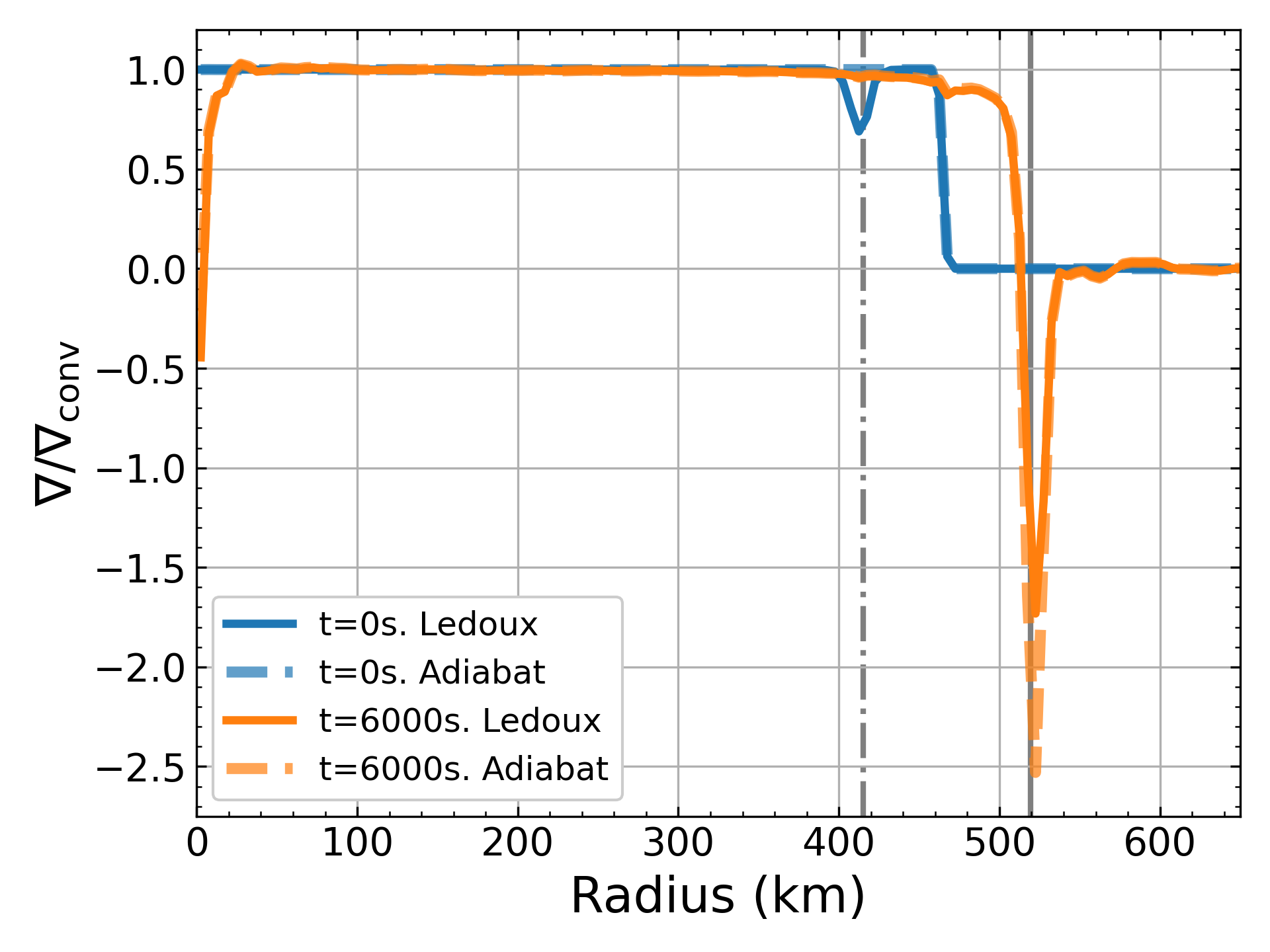}
    \caption{\label{fig:conv_grad} 
        Angle-averaged gradient ratios $\nabla / \nabla_{\mathrm{ad}}$ (dashed) and $\nabla / \nabla_{\mathrm{Led}}$ (solid) vs radius of the star.
        The blue curves represent the initial state and the orange curves represent the end of the high resolution simulation (i.e.\ after $6000 \, \second$ of simulation time). 
        The dot-dash grey vertical line indicates the location of the Urca shell, about $415 \, \mathrm{km}$. 
        The solid grey vertical line at about $520 \, \mathrm{km}$ marks \Rconv\ at the end of the simulation.
    }
\end{figure}

As the simulation evolves, the turbulent motions and mixing described by Figures \ref{fig:vrms_profiles} and \ref{fig:vort_profiles} lower the entropy of regions which were convectively stable, such that these regions become a part of the convection zone.
In 1D stellar modeling, the criterion for convective stability is generally defined by the Schwarzschild criterion:

\begin{equation}
    \frac{d \log T}{d \log P} < \left( \frac{d \log T}{d \log P} \right)_{\mathrm{ad}}
\end{equation}
where the ${}_\mathrm{ad}$ denotes the adiabatic gradient. Commonly, the gradients are denoted as $\nabla$ and $\nabla_{\mathrm{ad}}$ respectively.
When compositional gradients are important, it is necessary to add an additional term, $B$, to the right-hand side to compensate \citep{HKT2004}. 
$B$ can be defined by the gradient of the mean molecular weight, $\mu$:
\begin{equation}
    B = \frac{\chi_{\mu}}{\chi_T}  \frac{d \log \mu}{d \log P}
\end{equation}
where $\chi_{\mu} = \left( d \log P / d \log \mu \right)_{\rho, T}$ and  $\chi_{T} = \left( d \log P /d \log T \right)_{\rho, \mu}$ are thermodynamic derivatives. 
The addition of $B$ yields the Ledoux criterion:
\begin{equation}
    \frac{d \log T}{d \log P} < \left( \frac{d \log T}{d \log P} \right)_{\mathrm{ad}} + B
\end{equation}
where we denote the full right hand term as $\nabla_{\mathrm{Led}}$.

To calculate this compositional term in our simulations, we follow MESA's formulation (\citealt{paxton2013}, Equation 8), except we use centered difference and adjust for the spherical 3D geometry.
Here, $i,j,k$ are grid indices:
\begin{equation} \label{eqn:Bijk}
    B_{i,j,k} = -\frac{1}{\chi_T}\frac{\log P(\rho,T,X^{+}) - \log P(\rho,T,X^{-})}{\log P^{+}-\log P^{-}}
\end{equation}
where $\log P(\rho,T,X^{\pm})$ and $\log P^{\pm}$ are the sum of x, y, and z components. For example, the x components: 
\begin{equation}
    \log P(\rho,T,X^{\pm})_x = \frac{x}{r} \log P(\rho_{i,j,k},T_{i,j,k},X_{i \pm 1,j,k}) 
\end{equation}
\begin{equation}
    \log P^{\pm}_x = \frac{x}{r} \log P_{i \pm 1,j,k}
\end{equation}
The numerator in Equation \ref{eqn:Bijk} tracks the pressure gradient in response to changes in composition, with density and temperature remaining fixed. 
The denominator tracks the true pressure gradient.\footnote{The source code for this calculation is open-source and available at \url{https://github.com/AMReX-Astro/amrex-astro-diag}}

For our simulation, we calculate $\nabla$,  $\nabla_{\mathrm{ad}}$, and $\nabla_{\mathrm{Led}}$. 
The angle-averaged profiles are compared in Figure \ref{fig:conv_grad}, where we plot the ratio of the actual gradient, $\nabla$, to the two convective criterion gradients, $\nabla_{\mathrm{ad}}$ and $\nabla_{\mathrm{Led}}$.
According to the criteria, a ratio of less than one indicates a convectively stable region.
In the interior convection zone, we find the ratio is very close to one. 
This is expected as convection is very efficient and tends to drive $\nabla$ close to an adiabatic profile. 

The influence of the compositional gradients on $\nabla_{\mathrm{Led}}$ can be seen in the initial state (blue curves) where there is a dramatic spike at the Urca shell. 
However, as we evolve the simulation, the fluid is well mixed to the edge of the convection zone and the compositional contributions are largely negligible (see both orange curves are highly aligned), with an exception at the convective boundary where compositional gradients are quite large. 
Over time, the growth of the convection zone is displayed by the drop-off from a ratio of $\nabla/\nabla_{\mathrm{ad}} \sim 1$ moving radially outward, matching the \Xisot{C}{12} profiles in Figure \ref{fig:c12_profiles}. 
There is a large dip below zero just outside the convection zone (around $520 \, \mathrm{km}$), this is associated with the temperature valley seen in the orange curve of Figure \ref{fig:init_prof}. 
Here, the region is driven to lower temperatures as it follows the adiabatic curve, as compared to the initial isothermal profile.
The significance of this dip/temperature valley is discussed further in Section \ref{sec:discussion}.

\begin{figure*}
        \centering
            \includegraphics[width=0.95\textwidth]{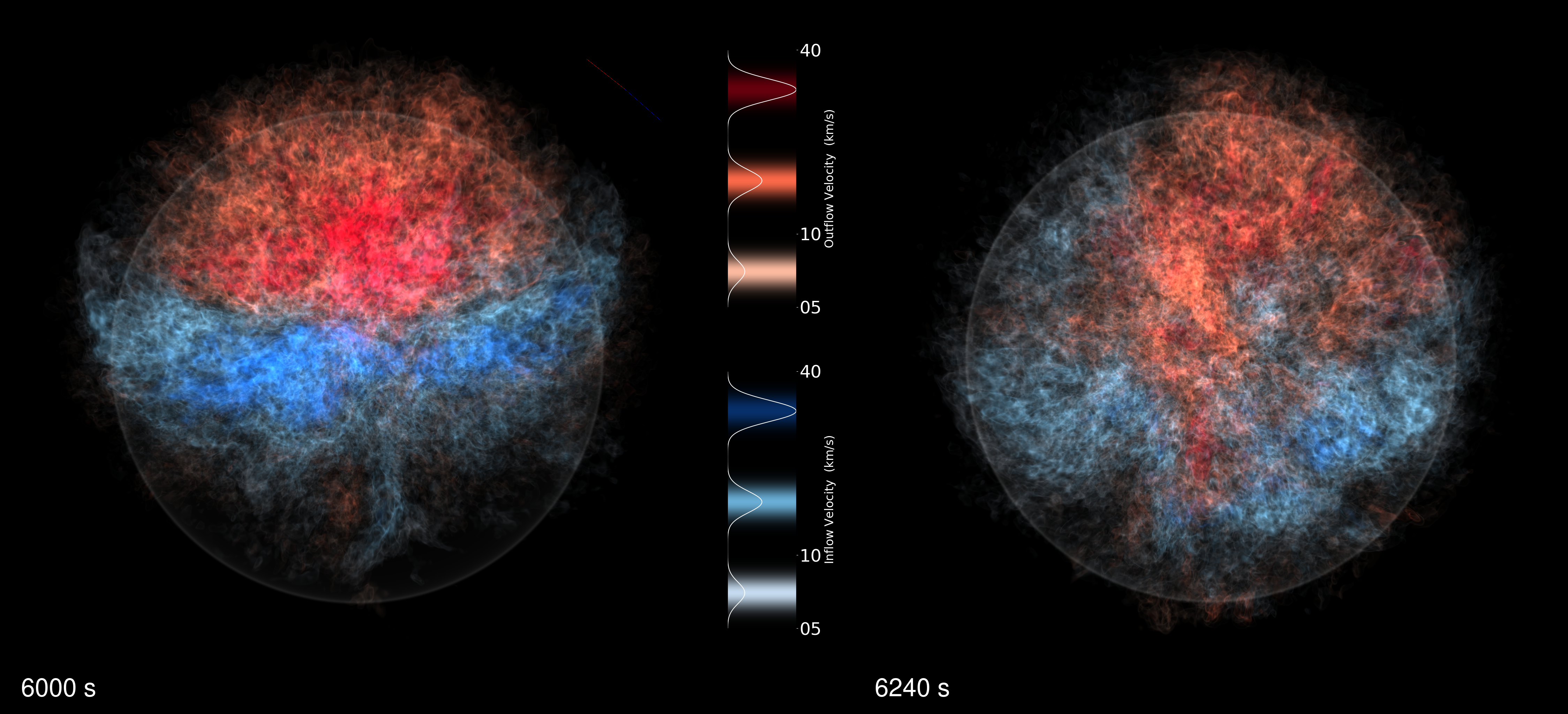}
    \caption{\label{fig:radvel} 
        Volume renders of the radial velocities in the convection zone for two snapshots of the high resolution simulation (see timestamp in lower left).
        Fluid moving radially outward from the center of the star are colored red.
        Fluid moving radially inward to the center of the star are colored blue.
        For each color, the shading indicates the magnitude, from darkest ($30 \, \kms$) to medium ($15 \, \kms$) to lightest ($7.5 \, \kms$).
        The white circle indicates the location of the Urca shell.
        The vertical axis of each volume render is oriented to align with the net radial flow (see Section \ref{subsec:vel_struct}).
        See HTML version of the article for an animation  which rotates about the center, displaying the full radial velocity structure. 
        The flow field appears to have a more ordered structure at 6000 s than at 6240 s.
    } 
\end{figure*}

\begin{figure*}
    \centering
    \includegraphics[width=\textwidth]{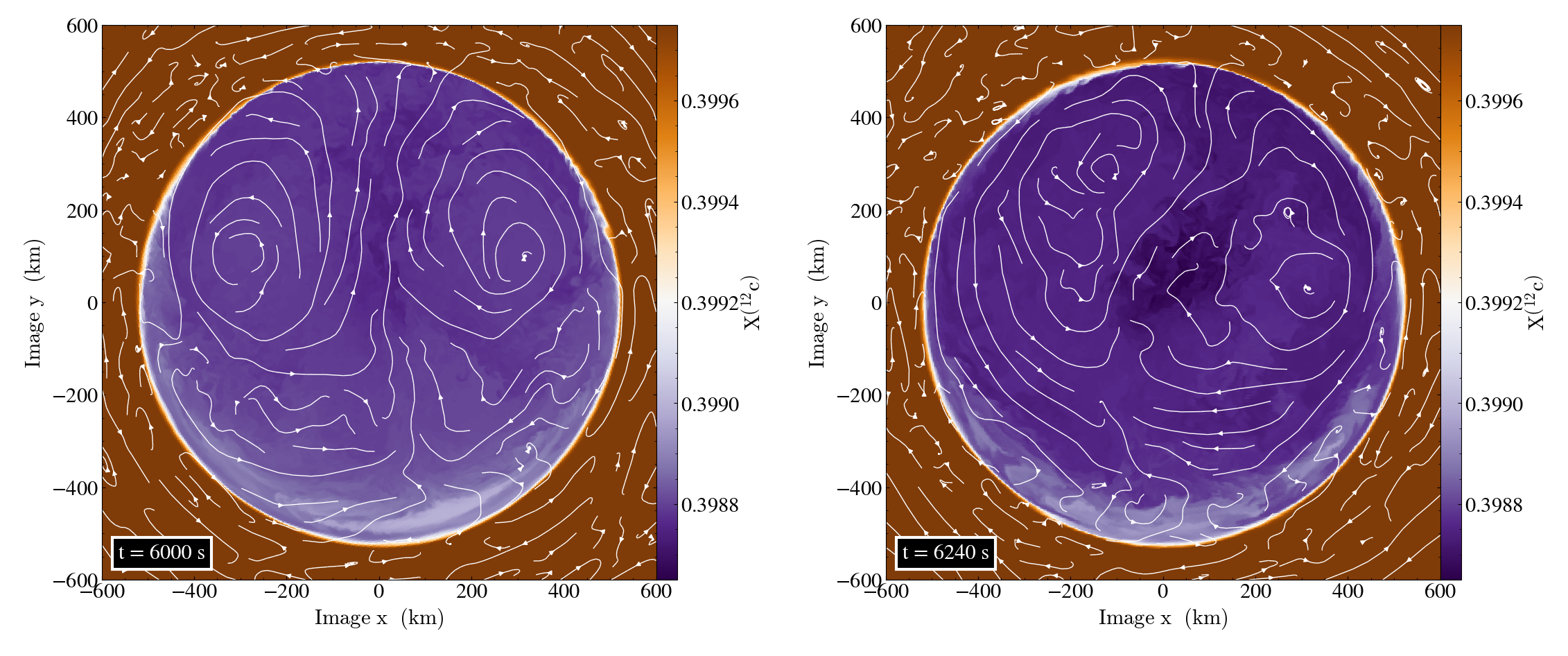}
    \caption{ \label{fig:c12_frac_sl}
        2D Slices through the center of the white dwarf, colored by the \Xisot{C}{12}, for two snapshots of the high resolution simulation (see timestamp in lower left).
        Streamlines are plotted on top in white, indicating the path a test particle would take through a given region of the star.
        The two slices are oriented as in Figure \ref{fig:radvel}. 
        So the vertical (``Image y") of each slice is oriented with the net radial flow as discussed in Section \ref{subsec:vel_struct}.
        }
\end{figure*}

\begin{figure}
    \centering
    \gridline{\fig{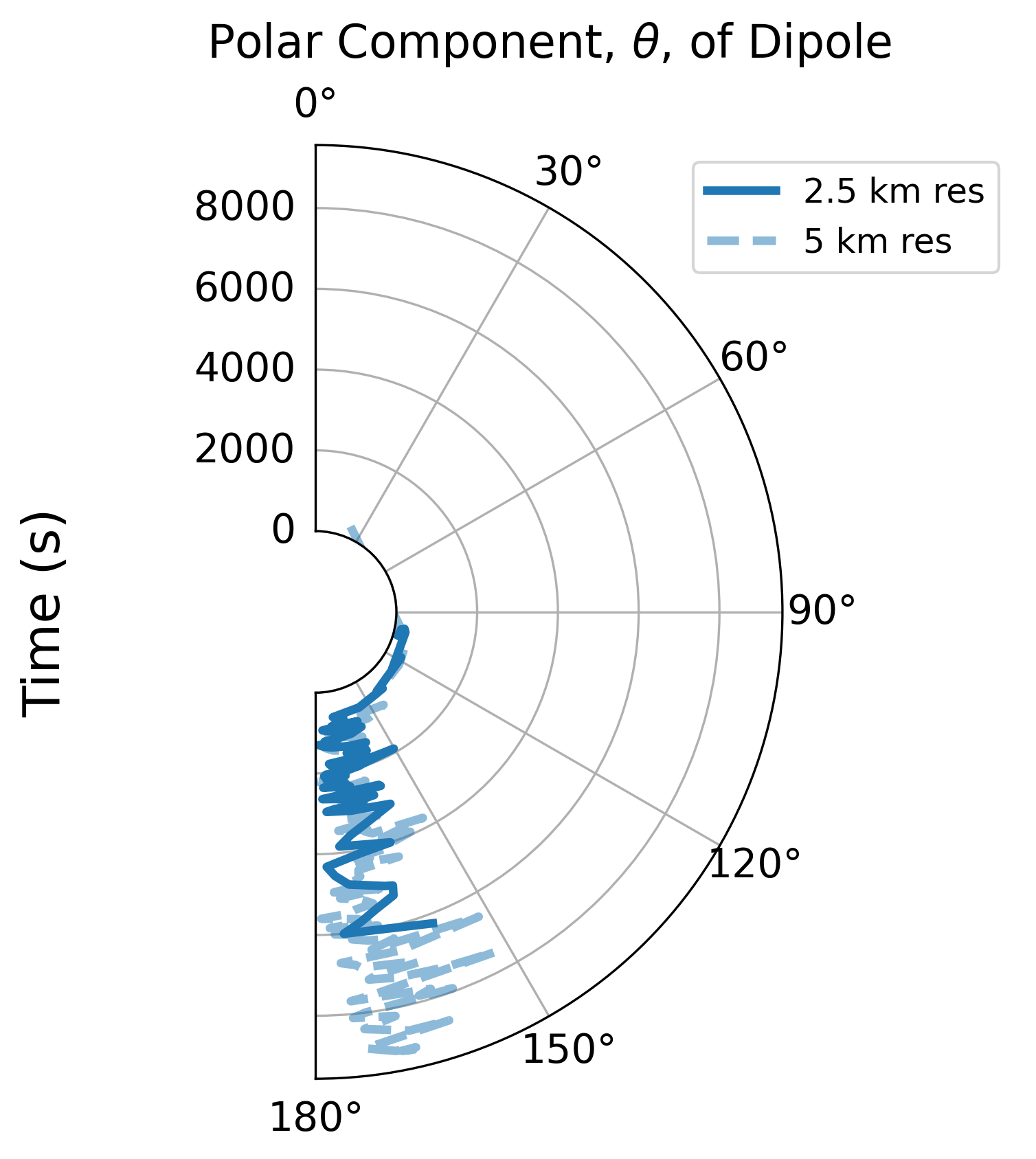}{0.8\linewidth}{(a)}}
    \gridline{\fig{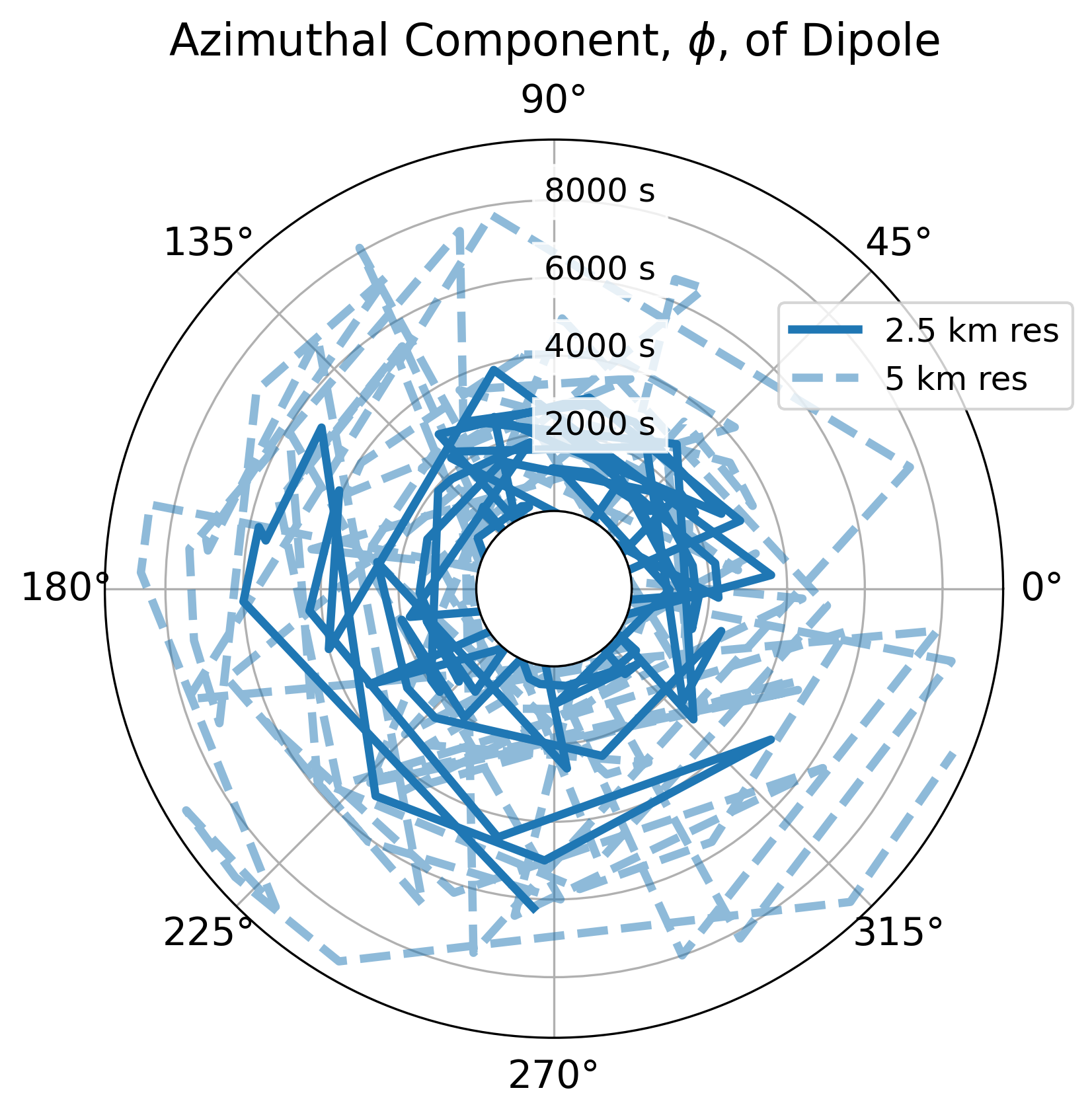}{0.8\linewidth}{(b)}}
    \caption{ \label{fig:dipole_angles}
        For each plot, the curves are plotted on a ``circular" grid to indicate the angle and the radial direction indicates time. 
        The solid curves represent the high resolution simulation (2.5 km) while the dashed curves represent the low resolution simulation (5 km).
        The top plot shows the polar angle, $\theta$, in relation to simulation time.
        $\theta$ is defined as the angle between the direction of net radial flow and the $z$-axis.
        The bottom figure plots the azimuthal angle, $\phi$, in relation to simulation time.
        $\phi$ is defined as the angle between the projection of the net radial flow onto the x-y plane and the $x$-axis.
        }
\end{figure}

\subsection{3D Velocity Structure} \label{subsec:vel_struct}

In the convection zone, the velocity field is dominated by a large-scale structure that extends to the outer regions of the convection zone (see left plot of Figure \ref{fig:radvel}). 
The velocity structure is ``fountain"-like, characterized by a large outflow of material from the center (red in Figure \ref{fig:radvel}) and with inflowing material more equatorially aligned (blue in Figure \ref{fig:radvel}) . 
By plotting streamlines on a slice through the simulation, as in Figure \ref{fig:c12_frac_sl}, we can directly see what appears to be a dipole structure that is noticeably off-center. 
It is the fact that this dipole is not aligned with the center of the star which creates the fountain-like appearance. 
However, the organized structure is not always present; as can be seen in the right plot of Figure \ref{fig:radvel}, the fountain structure breaks down on the timescale of several convective turnovers (${\sim} 200 \, \second$).
Throughout the simulation, the fountain structure regularly forms and dissipates, though there is generally some degree of dipole-like structure present (see the right plot of Figure \ref{fig:c12_frac_sl}). 
This behavior matches the intuition of a turbulent velocity field, which should regularly disrupt large scale structures.

The fountain structure has a clear directional dependence that shapes how material is mixed in the convection zone.
We calculate the direction of the radial flow by finding the density weighted averages of the radial velocity, $\langle U_r \rangle$, in the x, y, and z directions:

\begin{equation}
    \langle U_r \rangle_{x} = \left( \sum_{\mathrm{conv}}  \rho U_r \frac{x}{r} \right) / \sum_{\mathrm{conv}} \rho
\end{equation}
Here, we are summing over all cells in the convection zone.
$\langle U_r \rangle_{y}$ and $\langle U_r \rangle_{z}$ are calculated analogously. 
Using these values, we calculate the angle $\theta$ from the $z$-axis and the azimuthal angle $\phi$ (in the $x$-$y$ plane) as follows:

\begin{equation} \label{eqn:theta}
    \theta = \arctan \left( \frac{\sqrt{{\langle U_r \rangle_{x}}^2 + {\langle U_r \rangle_{y}}^2}}{\langle U_r \rangle_{z}} \right)
\end{equation}

\begin{equation} \label{eqn:phi}
    \phi = \arctan \left( \frac{\langle U_r \rangle_{y}}{\langle U_r \rangle_{x}} \right)
\end{equation}
Since the default nature of $\arctan$ is to span $[-\pi/2, \pi/2]$, Eqn. \ref{eqn:theta} is adjusted such that $\theta$ covers the proper range of $[0, \pi]$ with the zero point aligning with the ${+}z$-axis.
Similarly, we adjust Eqn. \ref{eqn:phi} based on the sign of $\langle U_r \rangle_{x}$ and $\langle U_r \rangle_{y}$ such that $\phi$ covers all angles, with the zero point aligning with the $x$-axis. 

In general, the direction of the flow is close to the $-z$ axis with a polar angle $\theta \approx 165 {}^\mathrm{o}$, see Figure \ref{fig:dipole_angles} (a). 
There are larger variations in the azimuthal angle, $\phi$, with time,  as seen in the Figure \ref{fig:dipole_angles} (b). 
The fact that the structure primarily points in a single coordinate axis throughout the simulation may be a numerical effect. 
It is unclear whether any other source could break the symmetry and favor the $-z$ axis in this way.
An additional explanation could be that once a direction is established from initial conditions, that axis will remain favored throughout the rest of the simulation.

The velocity structure alters how material is mixed throughout the convection zone. 
In Figure \ref{fig:c12_frac_sl}, we display the mass fraction of \isot{C}{12} in the convection zone.
Fluid from the center, which has a lower \Xisot{C}{12} (dark purple) due to carbon burning, is primarily transported to one side of the convection zone (top of Figure \ref{fig:c12_frac_sl}). 
This occurs whether the velocity field is highly structured (see left plot) or more chaotic (see right plot).
We see the impact most clearly on the opposite side, where there is an excess of \Xisot{C}{12} (light-purple to white) due to a lack of mixing.
This distribution of \Xisot{C}{12} demonstrates the non-isotropic mixing in the convection zone which is not well characterized by 1D stellar evolution models. 
The directionally dependent mixing will also have an impact on the convective Urca process and the distribution of the Urca pair in the simulation.

\begin{figure*}
    \centering
    \includegraphics[width=\textwidth]{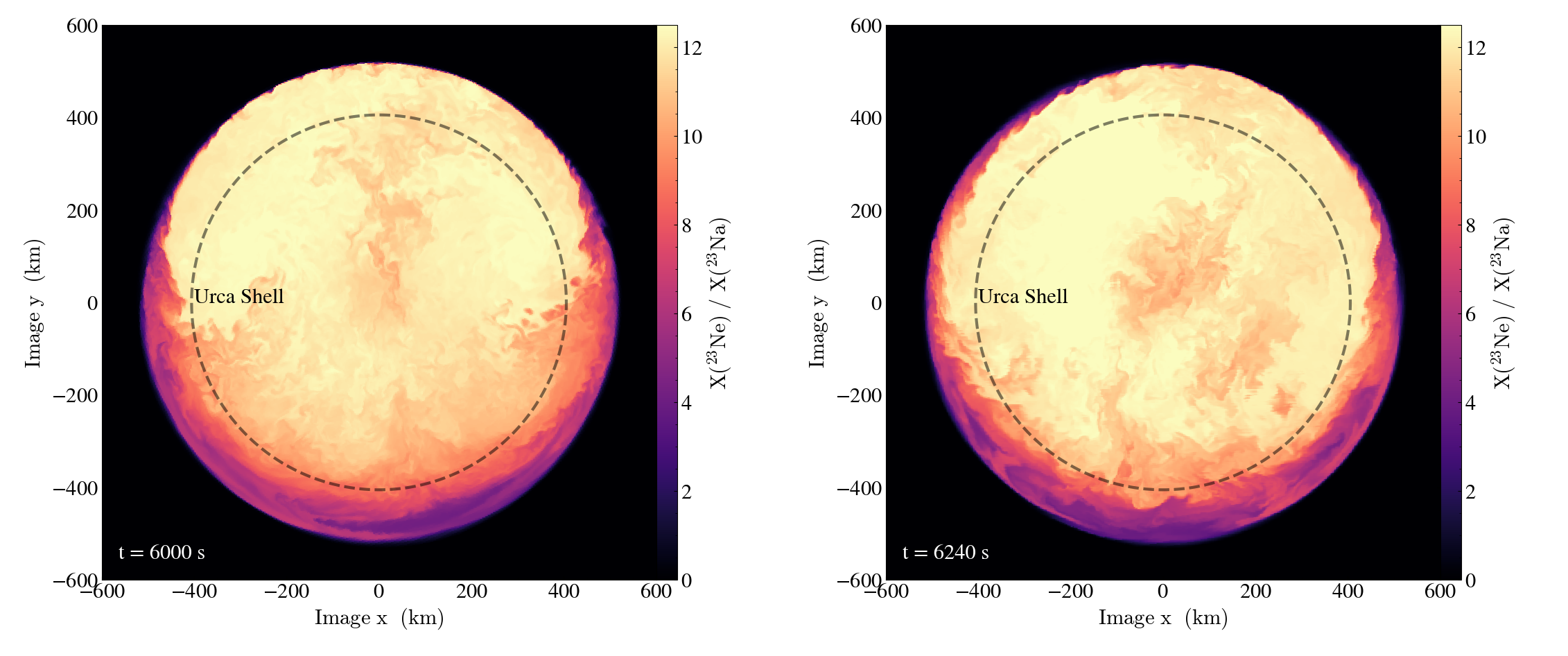}
    \caption{ \label{fig:urca_slice}
        2D slices through the center of the white dwarf, colored by the ratio of $\Xisotm{Ne}{23}/\Xisotm{Na}{23}$, for two snapshots of the high resolution simulation (see timestamp in lower left).
        The dashed circle indicates the location of the Urca shell.
        The two slices are oriented as in Figure \ref{fig:radvel}, 
        such that the vertical (``Image y") of each slice is oriented with the net radial flow as discussed in Section \ref{subsec:vel_struct}.
    }
\end{figure*}

\begin{figure}
    \centering
    \plotone{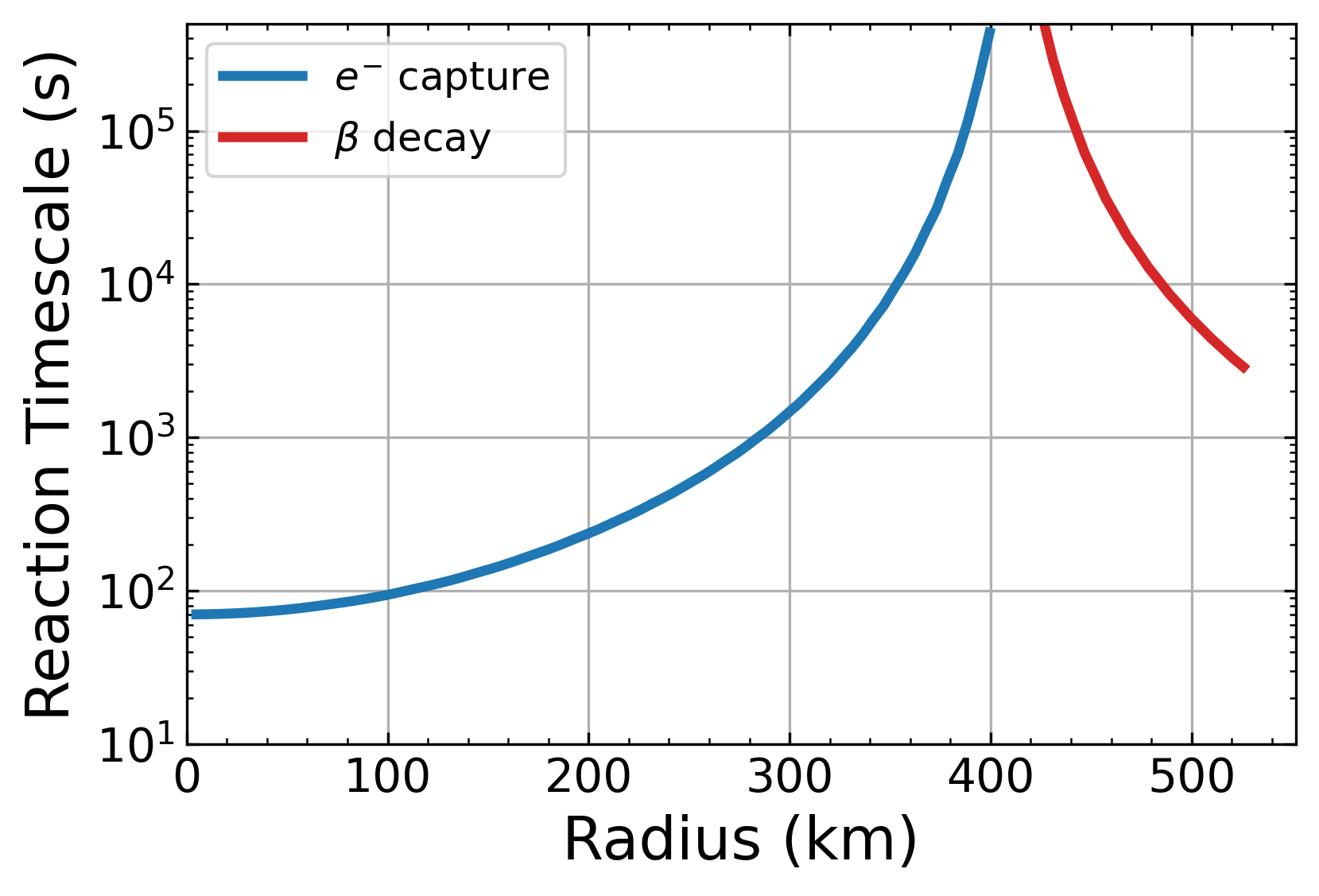}
    \caption{ \label{fig:urca_timescale}
        The timescales associated with the Urca reactions vs radius of the star.
        The blue curve represents electron capture timescale, \Tauecap, interior to the Urca shell.
        The red curve represents $\beta$-decay timescale, \Taubeta, outside the Urca shell.
        These timescales were calculated using the end of the high resolution simulation (i.e.\ after $6000 \, \second$ of simulation time). 
    }
\end{figure}

\begin{figure}
    \centering
    \plotone{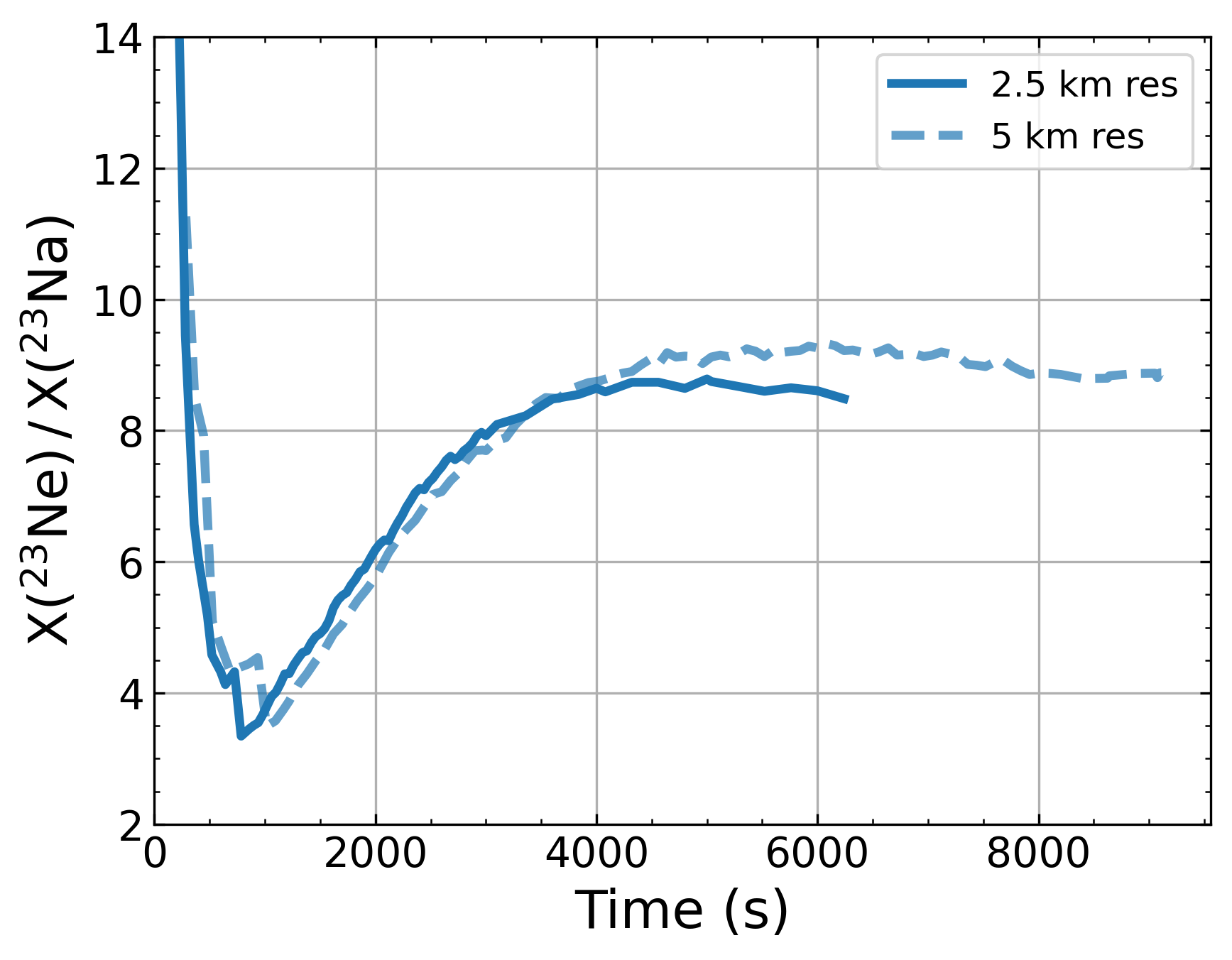}
    \caption{ \label{fig:ratio_overtime}
        Time variation of the average ratio $\Xisotm{Ne}{23} / \Xisotm{Na}{23}$ in the convection zone.
        The solid curve represents the high resolution simulation (2.5 km) while the dashed curve represents the low resolution simulation (5 km).
    }
\end{figure}

\begin{figure*}
    \centering
    \includegraphics[width=\textwidth]{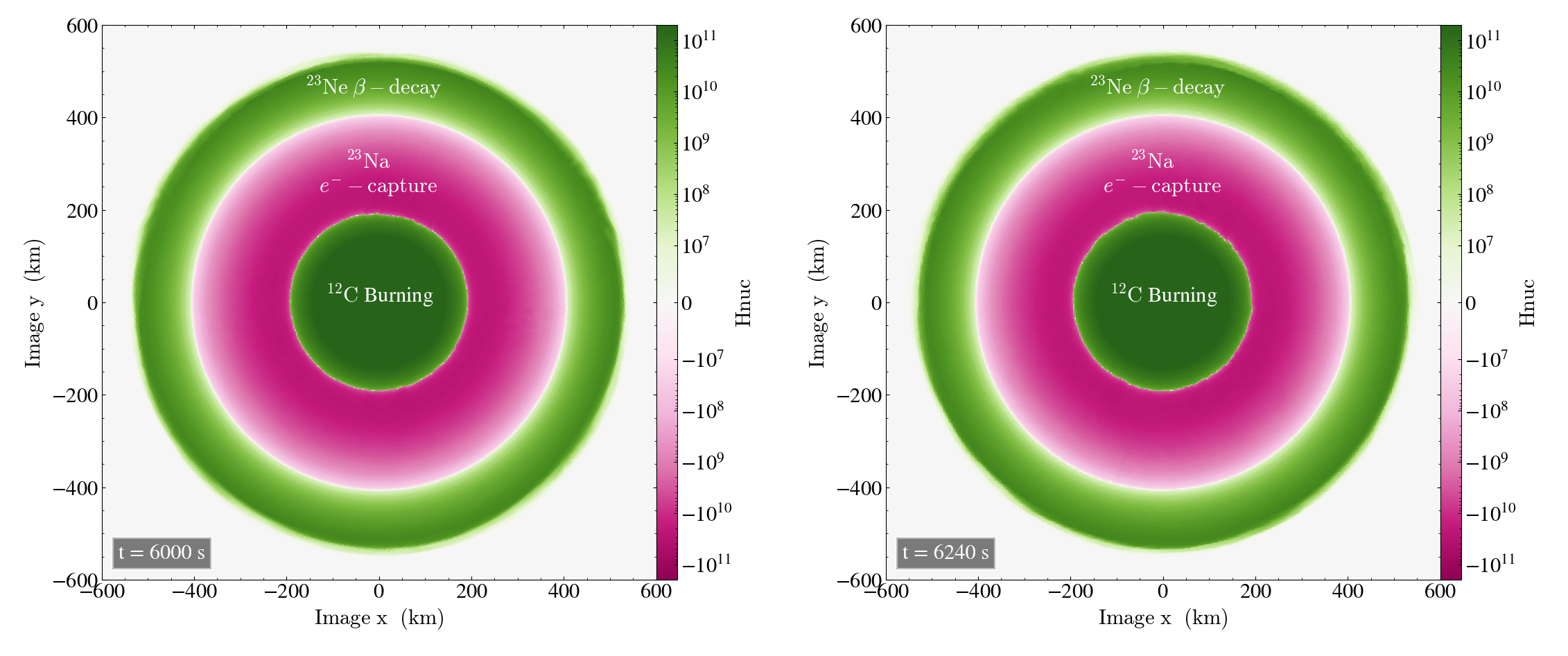}
    \caption{\label{fig:hnuc} 
        2D slices through the center of the white dwarf, colored by the specific energy generation rate (in $\erg~\unitstyle{g}^{-1}~\second^{-1}$), for two snapshots of the high resolution simulation (see timestamp in lower left).
        The dominant (i.e.\ the most active) reaction in each region is annotated in white, as well as the location of the Urca shell.
        The two slices are oriented as in Figure \ref{fig:radvel}, 
        such that the vertical (``Image y") of each slice is oriented with the net radial flow as discussed in Section \ref{subsec:vel_struct}.
    }
    
\end{figure*}

\begin{figure}
    \centering
    \plotone{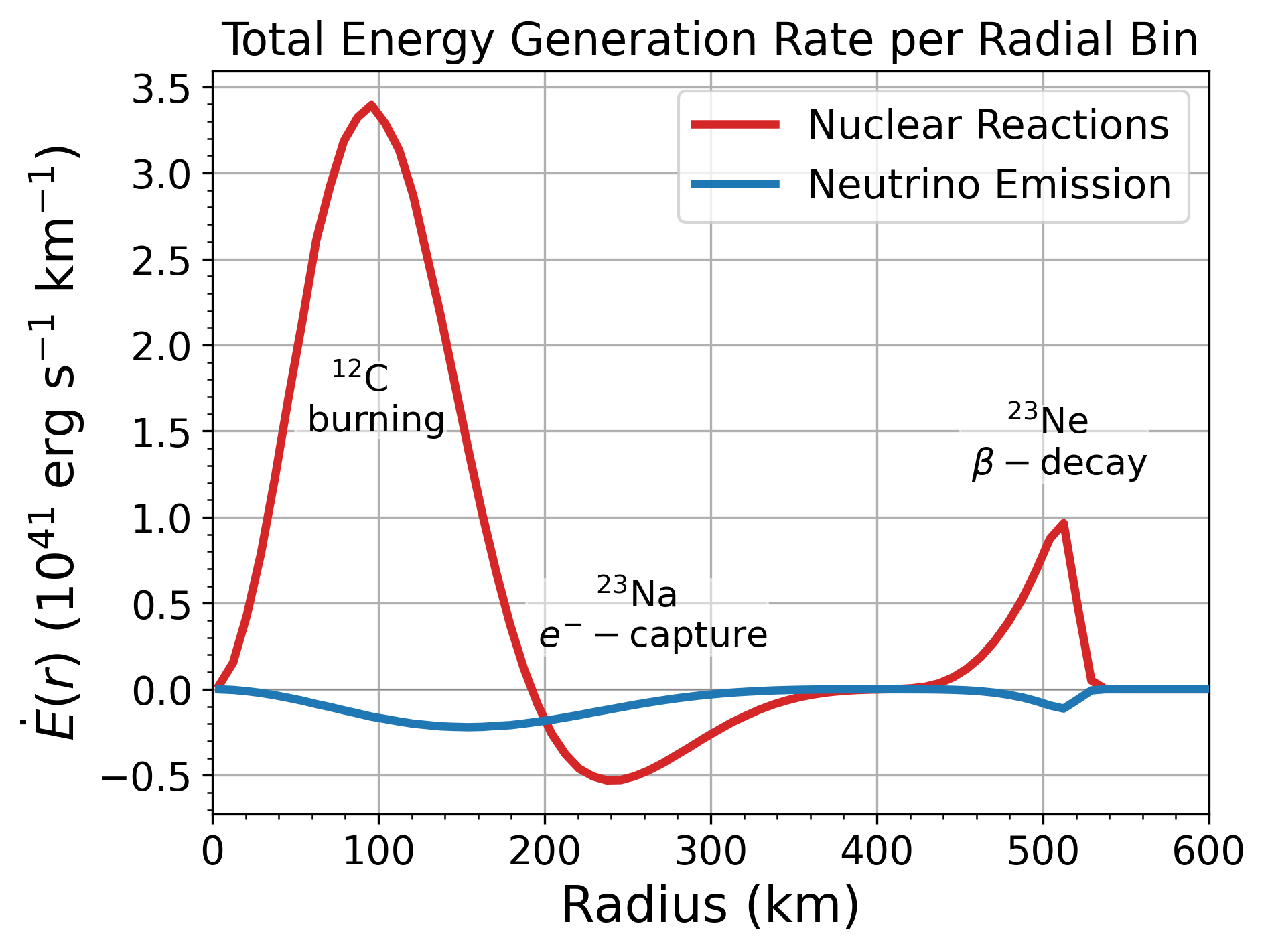}
    \caption{ \label{fig:nu_energy_loss}
        The angle-averaged energy generation/loss rate per radial bin vs radius of the star.
        The red curve shows energy generated by all nuclear reactions.
        The blue curve shows the energy losses due to neutrino emission. 
        The dominant (i.e.\ the most active) reaction in each region is annotated in black.
    }
\end{figure}

\subsection{Urca Process} \label{subsec:urca_proc}

Similar to \Xisot{C}{12}, the distribution of the Urca pair is highly aligned with the velocity structure (see Figure \ref{fig:urca_slice}).
The large outflow of fluid from the center brings \isot{Ne}{23} rich material in the core, out past the Urca shell and to the edge of the convection zone. 
Once outside the Urca shell, the \isot{Ne}{23} slowly $\beta$-decays into \isot{Na}{23}. 
We estimate the $\beta$-decay timescale, \Taubeta, outside the Urca shell and the electron-capture timescale, \Tauecap, inside the Urca shell as:

\begin{equation}
    \Taubetam \sim \frac{\Xisotm{Ne}{23}}{\dot{\omega}_{\isotm{Na}{23}}}
\end{equation}

\begin{equation}
    \Tauecapm \sim \frac{\Xisotm{Na}{23}}{\dot{\omega}_{\isotm{Ne}{23}}}
\end{equation}

These two timescales vary with density which can be seen in their angle-averaged profile through the convection zone, shown in Figure \ref{fig:urca_timescale}.
The shortest timescale, $\Tauecapm \sim 100 \, \second$, occurs at the center of the star where densities are highest. 
This puts \Tauecap\ roughly on the order of \Tauconv\ for the inner $100 \, \unitstyle{km}$ in radius. 
For larger radii, around $300$-$500 \, \unitstyle{km}$, both Urca timescales are at least an order of magnitude longer than \Tauconv.
This large difference in timescales peaks near the Urca shell, around $415 \, \unitstyle{km}$ in radius, as the two Urca timescales quickly balloon.

The relatively short timescale of \Tauconv\ in much of the convection zone can be seen in Figure \ref{fig:urca_slice}, where there remains a large fraction of \isot{Ne}{23} outside the Urca shell.
The mixing of \isot{Ne}{23}-rich material across the shell occurs significantly faster than the $\beta$-decay of \isot{Ne}{23}, particularly for regions dominated by the large outflows like in the top portion of Figure \ref{fig:urca_slice}.
It is only in the regions with relatively less mixing from the core, see the bottom purple portion of Figure \ref{fig:urca_slice}, that the \isot{Ne}{23} has time to $\beta$-decay and we see appreciable amounts of \isot{Na}{23} in the convection zone.

We compare the result of our simulations to that of the ``quick mixing" limit. 
In this limit, the mixing is efficient enough to maintain uniform composition throughout the convection zone. 
With this assumption, we calculate the predicted equilibrium ratio of the Urca pair, $\Xisotm{Ne}{23}/\Xisotm{Na}{23}$, by integrating the reaction rates over the volume of the convection zone. 
For our simulation, the result of the quick mixing limit is $\Xisotm{Ne}{23}/\Xisotm{Na}{23} \approx 46$. 
Note however, this rough estimate does not account for the sources of \isot{Na}{23} from the carbon burning or from the growth of the convection zone.
The quick mixing limit is only used as a point of reference.
In comparison to this limit, we find a much lower ratio of $\Xisotm{Ne}{23}/\Xisotm{Na}{23} \sim 9$ as shown in Figure \ref{fig:ratio_overtime}. 
The smaller ratio indicates a greater portion of \isot{Na}{23} is present in our simulation due to the sources of \isot{Na}{23} stated previously and the slow non-uniform levels of mixing. 
The non-isotropic nature of the mixing, which results in the purple areas of Figure \ref{fig:urca_slice}, greatly lower the average ratio of $\Xisotm{Ne}{23}/\Xisotm{Na}{23}$.

We further analyze the sources of \isot{Na}{23} and thus the overall amount of Urca pair in the convection zone by calculating the mass generation rate of \isot{Na}{23}.
At the end of the simulation run, the rate of \isot{Na}{23} mass generated by carbon burning is $1.30 \times 10^{-4} \, \Msun~\hour^{-1}$.
Additionally, the growth of the convection zone, as seen in Section \ref{subsec:conv_zone}, leads to further \isot{Na}{23} rich material being added to the convection zone.
Using the initial mass fraction of $\Xisotm{Na}{23} = 5 \times 10^{-4}$ and our estimated $\dot{M}_{\mathrm{conv}}$, we calculate the rate at which \isot{Na}{23} is added due to the growing convection zone to be ${\sim}2.5 \times 10^{-5} \, \Msun~\hour^{-1}$.
In total, these two sources roughly double the amount of Urca pair in the convection zone from an initial $2.5 \times 10^{-4} \, \Msun$ to about $5 \times 10^{-4} \, \Msun$ at the end of the simulation. 

The convection zone is split into three distinct regions where a single nuclear reaction is of primary importance as annotated in Figure \ref{fig:hnuc}. 
In the center, carbon burning produces a large amount of energy. 
Further out in radius, the carbon burning rate drops and instead the electron capture of \isot{Na}{23} to \isot{Ne}{23} dominates. 
Note, this is an endothermic reaction which explains the negative energy generation rate (pink in Figure \ref{fig:hnuc}). 
Around a radius of about $415 \, \unitstyle{km}$ lies the Urca shell which separates the pink electron capture region from the green $\beta$-decay region on the outskirts of the convection zone.

As seen in Figure \ref{fig:hnuc}, the energy generation rate is quite radially symmetric.
This is due to the strong density dependence of the nuclear reaction rates. 
Through the changing density of the convection zone, the rates vary by several orders of magnitude (see range in Figure \ref{fig:urca_timescale}), while the mass fractions of the Urca pair only vary by a factor of 12 (see range in Figure \ref{fig:urca_slice}).
In Figure \ref{fig:nu_energy_loss}, we compare the rate of energy generated by all the nuclear reactions to the rate of energy lost to free-streaming neutrinos. 
The neutrino emissions come from the Urca reactions as well as thermal emission from the plasma.
Summing over the total volume, we find the total energy generated by nuclear reactions to be $\dot{E}_{\mathrm{nuc}} = 3.32 \times 10^{43} \; \erg~\second^{-1}$, dominated primarily by the carbon burning near the center (see the annotated regions in Figure \ref{fig:hnuc} and Figure \ref{fig:nu_energy_loss}). 
In comparison, the total energy lost to neutrinos is about an order magnitude less, $\dot{E}_{\nu_e {-} \mathrm{tot}} = 4.18 \times 10^{42} \; \erg~\second^{-1}$.

Note that a significant portion of the neutrino losses directly follow carbon burning.
This is because \isot{Na}{23} is a product of this burning, and quickly reacts to \isot{Ne}{23} near the center of the star, emitting a significant amount of neutrinos.
To estimate the energy losses due to these neutrinos, we compared the rate of \isot{Na}{23} generated solely by carbon burning to the rate of \isot{Na}{23} destroyed by electron captures. 
We found the ratio of these two generation rates to be about 0.73. 
Assuming the \isot{Na}{23} generated by carbon burning is consumed locally, this indicates that about 70\% of the neutrino losses from electron capture reactions follow from the carbon burning.
Using this ratio, we estimate that about 66\% ($2.77 \times 10^{42} \; \erg~\second^{-1}$) of the total neutrino energy losses are a result of carbon burning, not the convective Urca process. 
In contrast, the cooling by thermal neutrino emission is marginal with $\dot{E}_{\nu_e {-} \mathrm{thermal}} = 4.44 \times 10^{38} \; \erg~\second^{-1}$.
After accounting for these losses, we find the energy lost directly due to the convective Urca process is about $1.41 \times 10^{42} \; \erg~\second^{-1}$.
In total, though the neutrinos take energy out of the star, the losses do not outweigh the energy generated by carbon burning. 
Summing the contribution from all neutrino emission adds up to losses of 12.6\%, with an estimated 4.2\% resulting from the convective Urca process.

These neutrino losses and general nuclear energy generation are related to the amount of carbon burned, $\Delta M_{\isotm{C}{12}}$, which impacts the neutron excess or neutronization, $\eta = \sum_i X_i/A_i (A_i - 2 Z_i)$ where $A$ is the mass number, $Z$ is the proton number and the sum is over all isotopes.
After the transitory period, around $t = 3000 \, \mathrm{s}$, we calculate $\Delta M_{\isotm{C}{12}} = 3.14 \times 10^{-4} \, \Msun$ and $\Delta \bar{\eta} = 4.13 \times 10^{-5}$ over the quasi-steady state portion of the simulation.
Here, $\bar{\eta}$ is the mass weighted average of $\eta$ over all cells in the convection zone.
These results can be compared to previous estimates \citep{piro2008, chamulak2008} and results from 1D stellar evolution models of the simmering phase \citep{martinez-rodriguez2016, schwab2017a, piersanti2017, piersanti2022}.
This comparison however is non-trivial due to the previous results spanning the full simmering phase, while the presented simulations only capture a roughly one hour snapshot of the burning.
Additionally, the various estimates and models make different assumptions on the reaction network and the role of the Urca process.
In our presented simulations, we leave out important rates for capturing the proper change in neutronization, particularly $\isotm{C}{12}(p,\gamma)\isotm{N}{13}$ and subsequent electron capture.
We discuss this further in Section \ref{sec:conclusion} as noted improvements we hope to include in future work.

\section{Discussion} \label{sec:discussion}
In the presented simulations, we found a growing convection zone driven by carbon burning at the center.
This is despite the  simulations reaching a quasi-steady state, evidenced by the \vrms\ and $\Xisotm{Ne}{23}/\Xisotm{Na}{23}$ relaxing to constant values over many convective turnovers (Figure \ref{fig:vrms_conv} and \ref{fig:ratio_overtime} respectively).
Even after these values had reached an apparent equilibrium, the convection zone continued to expand.
This expansion of the convection zone resulted in the formation of a  temperature valley as the adiabatic profile extended into the stable isothermal region, as seen in Figure \ref{fig:init_prof}. 
We interpret this temperature valley as the realistic temperature structure of the convective boundary region for a simmering white dwarf. 
Unlike a core convection zone inside a long-lived star, the energy deposited at the center during the simmering phase only contributes to the heating and growth of the convection zone.
The thermal conduction times of the overlying layers are too long for significant amounts of energy to be transported out of the convection zone during this short simmering phase.

Inspecting our final state suggests the temperature valley is a result of the convective boundary mixing. 
For any large outflow, the deceleration region (see Section \ref{subsec:vel_struct}) is likely to extend close to the convective boundary predicted by Mixing Length Theory (MLT). 
That is the intersection of the inner adiabatic profile of the convection zone and the temperature profile of the convectively stable outer envelope.
Thus, there is expected to be a well-mixed, and therefore nearly adiabatic, region that extends beyond the MLT predicted edge of the convection zone.
In the context of the simmering phase, where thermal timescales are long, this well-mixed region would produce the observed temperature valley.
The size of this well-mixed region is expected to be related to both the deceleration region as well as the outward diffusion of the turbulence field, generated by the core convection. 
Although more investigation is required before a conclusion can be drawn, it appears plausible that this velocity structure, and not the growth rate, is the main determining factor for the size of this well-mixed region and thus the temperature valley.

In addition to the velocity structure, the abundances involved in the Urca process would also appear to be important.
As mentioned in Section \ref{sec:urca}, the Urca process alters the electron fraction gradient, which should effect the structure of the deceleration region, and possibly the diffusion of turbulence into overlying layers.
As seen in Figure \ref{fig:conv_grad}, the departure of the temperature gradient from the adiabatic profile appears to be impacted by the presence of the Urca shell.

This quickly growing convection zone, as compared to the thermal timescale, and convective boundary mixing presents a situation that leads to a fundamental shortcoming of the classical MLT.
As the convection zone expands into the isothermal regions in the initial model, it must ingest material with higher entropy than the current convection zone. 
Thus, the \textit{inward} ingestion of material should lead to a net transfer of energy into the convection zone.
This, however, runs counter to classical MLT, which by construction, can only transfer energy \textit{outward}. 
Typical prescriptions of convective boundary mixing for 1D stellar evolution, such as those implemented in MESA, do not address this issue. 
These prescriptions are mainly concerned with mixing of composition, leaving the heat transfer to follow classical MLT.

We do not undertake the construction of a proposed quasi-steady state temperature structure that includes the valley due to mixing from the expanding convection zone, leaving that to future work.
However, such a consistent initial condition should allow a much faster approach to equilibrium since the velocity field itself develops fairly quickly, see Figure \ref{fig:vrms_conv}.

We do note that a somewhat similar feature is found in the 1D stellar models presented in Figures 8 and 11 of \cite{piersanti2022}.
These figures show the temperature profile for various 1D models, and indicate a dip at the edge of the convection zone.
However, it is unclear if this is the same feature as the temperature valley seen in Figure \ref{fig:init_prof}.
The profiles presented in \cite{piersanti2022} are of the final moments of the simmering phase (near $T_{\mathrm{c}} \approx 8 \times 10^8 \, \mathrm{K}$) and the dips are accompanied by a very large drop in temperature from the convective core to the isothermal exterior.

We found the convection zone to be dominated by a large scale fountain-like velocity structure, resulting from an off-center dipole structure, see streamlines on Figure \ref{fig:c12_frac_sl}.
Though the velocity structure was intermittently disrupted by the turbulent flow, the general picture remained of highly non-isotropic convective mixing.
This structure of mixing is not well described by a local diffusive law, like that derived from MLT.
MLT does not capture the the non-local behavior of the large structure, particularly the radially extending flows that span from the center of the star to the edge of the convection zone.
To properly model this mixing behavior in a 1D stellar evolution simulation, a non-local prescription may be necessary.
The stable direction of the dipole moment in our simulations (as seen in Figure \ref{fig:dipole_angles}) further exacerbates this non-locality issue.
But even if this is taken as a simple numerical effect, the large radial structure remains an issue for local mixing prescriptions.

The convective mixing is the dominant driver of the distribution of the Urca pair. 
This result is most clearly seen in the radial profiles of \isot{Ne}{23} and \isot{Na}{23} (see of Figure \ref{fig:init_prof}).
The transition between the Urca pair occurs at the convective boundary, not the Urca shell.
However, the distribution of the Urca pair is far from uniform as shown in Figure \ref{fig:urca_slice}. 
This is due to the mixing rate and reaction rates varying throughout the convection zone. 
The non-uniformity of the mixing rate is best seen in Figure \ref{fig:c12_frac_sl}. 
Here, \isot{C}{12} is unevenly distributed, particularly in the outer regions of the convection zone. 
The streamlines plotted on Figure \ref{fig:c12_frac_sl} indicate the large eddies which primarily mix regions in the top half of the plot, without extending to the regions near the bottom of the plot.
This leads to an excess of \isot{C}{12} (see the light-purple/white regions near the bottom of the plot). 

In addition to the non-uniform mixing, the Urca reactions also vary throughout the convection zone as demonstrated in Figure \ref{fig:urca_timescale}. 
Near the center, inside $100 \, \mathrm{km}$ in radius, the electron capture rate is actually comparable to the mixing rate, $\Tauecapm < 100 \, \mathrm{s}$.
This results in \isot{Na}{23} that is produced by carbon burning quickly reacting to \isot{Ne}{23}.
However, the Urca reactions are much slower in most other regions of the convection zone, leading to the mixing playing a more dominant role.
Thus we find the Urca distribution to be very similar to the \isot{C}{12} distribution as seen in Figures \ref{fig:c12_frac_sl} and \ref{fig:urca_slice}.

Despite some regions, namely near the center, where the Urca reactions were on similar timescales to the mixing timescale, we did not find that the convective Urca process hinders the extent of the convection zone to the Urca shell as seen in \cite{stein-wheeler2006}.
Instead, the convection zone grows well past the Urca shell as seen in Figure \ref{fig:conv_mass}, and was not able to reach a stable size during the simulation run. 
As stated in Section \ref{subsec:conv_zone}, the stable size of the convection zone may be much larger and align more closely to the fiducial model in \cite{martinez-rodriguez2016}. 
However, that model also found the convectively stable overlying regions to be at temperatures a little over $1 \times 10^8 \, \mathrm{K}$. 
From our parameterized initial model, we set a much hotter isothermal region of $3 \times 10^8 \, \mathrm{K}$.
Additionally, the 1D model did not account for the direct impact of the convective Urca process on convection or include the temperature valley that may be characteristic of the simmering phase.

Broad conclusions on the effects of the Urca process on convection should not be determined strictly from this simulation. 
This simulation covers only a brief period of time, about a few hours, in the evolution of a simmering white dwarf, which spans thousands of years. 
More work is needed to better constrain the effects of the convective Urca process on convection. 
This includes for varied initial conditions that represent different stages of the simmering phase.
During these stages, the mixing and reaction timescales will vary and may have large impacts on the importance of the convective Urca process to the white dwarf.
Additionally, as the convection zone did not reach a stable size in this simulation, it is possible that the convective Urca process still has some hindering effects.

\section{Conclusion and Future Work} \label{sec:conclusion}
The convective Urca process is an important piece to understanding a simmering white dwarf as a Type Ia SNe progenitor.
Through the use of the \maestro\ low Mach number code, we were able to perform the requisite long-time-scale simulations, on order a hundred convective turnovers, of the convection zone of a simmering white dwarf. 
These simulations allowed us to characterize the evolution of the system and the coupling of turbulent convection and weak Urca reactions.
The simulations indicated the convective mixing had a strong directional dependence which influenced the distribution of the Urca pair, \isot{Na}{23} -- \isot{Ne}{23}. 
The highly non-local structure of this mixing is not well characterized by a local diffusive law like MLT.
While our simulations reach equilibrium with respect to the magnitude of the convective velocity and balance of Urca reactions, the mixed region was still quickly expanding after hundreds of convective turnover time.
One of the potential causes for this is our initial choice of a $0.5 \, \Msun$ convection zone was too small. 
The fact that a larger convection zone would be more stable disagrees with the idea that the convective Urca process restricts convection to the Urca shell.
In our simulations, we found the energy lost to neutrinos did not outweigh the energy released by carbon burning, and thus no net cooling of the star occurred.
Of the total power generated by nuclear reactions, 12.6\%  is lost to neutrino emission, with an estimated third of those losses coming from the convective Urca process, two thirds from the immediate reaction of \isot{Na}{23} produced by carbon burning, and a negligible amount from thermally emitted neutrinos.

These simulations indicate that the temperature structure of the outer boundary of the convection zone is qualitatively different than that inferred from MLT, even with consideration of overshooting.
We have found that the expanding convection zone creates a temperature minimum just outside the location where standard criteria would predict the convective boundary.
It appears that this region is related to the turbulent mixing of material at the boundary into the convection zone, thus lowering its entropy with respect to the overlying material.
This mixed region extends the near-adiabatic profile down beyond the temperature of the overlying material.
Abundances of Urca pairs in this region indicate that its structure is likely influenced by the convective Urca process.
Unfortunately, we postpone further characterization of this novel convective boundary layer structure to future work since we were not able to fully reach steady state in this work.
It is expected that further improvements to the initial condition may put this within reach.

Further simulations will be needed to properly investigate the extent to which the convective Urca process impacts the simmering white dwarf.
In particular, comparing simulations with and without the convective Urca process (similar to the methods in \citealt{stein-wheeler2006}) should demonstrate the extent to which the Urca process cooling dampens or hinders the convection in the white dwarf.
Additionally, the use of more comprehensive reaction networks, which include more accurate prescriptions of neutron decay \citep{langanke2001} and additional carbon burning related rates particularly $p$ and $n$ capture rates), will ensure the carbon burning driven convection is properly modeled.
Furthermore, adding additional Urca pairs such as \isot{Ne}{21} -- \isot{F}{21} and \isot{Mg}{25} -- \isot{Na}{25} to our reaction network will help in capturing a more complete picture of the Urca cooling in our simulations.
In addition to expanded reaction networks, we look to explore the use of an explicit integrator scheme to accelerate the nuclear integration in our simulations.
We are also interested in using different central conditions to investigate the impact of the convective Urca process on different stages of simmering, as well as different progenitors produced from varying model parameters (i.e.\ higher or lower accretion rates). 
This includes earlier times in the simmering, when the Urca shell is located just exterior to the MLT predicted convective boundary (i.e.\ in an "overshooting" region), as well as progenitor models with lower densities (as low as $\rho_c \sim 1{-}1.5 \times 10^9 \, \gcc$) and a range of central temperatures to correspond with the approach to the ignition of the flame (around $T_c = 8 \times 10^8 \, \mathrm{K}$).

\begin{acknowledgments}
The authors acknowledge the earlier dissertation work of Don Willcox \citep{willcox2018} on which this work builds.
The authors thank Catherine Feldman for fruitful conversations.
\maestro\ is freely available on GitHub (\url{https://github.com/AMReX-Astro/}), and all problem setup files for the calculations presented here are in the code repository.  
This research was supported in part by the US Department of Energy (DOE) under grant DE-FG02-87ER40317.
The reaction networks were generated using the \pynucastro\ library \cite{smith2023}.
This research used resources of the National Energy Research Scientific Computing Center, which is supported by the Office of Science of the U.S. Department of Energy under Contract No. DE-AC02-05CH11231, using NERSC award NP-ERCAP0027167.
Visualizations and part of this analysis made use of \yt\ \cite{turk2011}. 

\end{acknowledgments}

\software{\amrex\ \citep{zhang2019},                 
          {\sffamily matplotlib}\  \citep{Hunter:2007},
          \maestro\ \citep{fan2019,maestroex_joss},
          {\sffamily NumPy}\ \citep{numpy2020},
          \pynucastro\ \citep{pynucastro,smith2023},
          {\sffamily SymPy} \citep{sympy}, yt \citep{turk2011}}
         
\facilities{NERSC}

\bibliography{urca.bib}{}

\begin{thebibliography}{}
\expandafter\ifx\csname natexlab\endcsname\relax\def\natexlab#1{#1}\fi
\providecommand{\url}[1]{\href{#1}{#1}}
\providecommand{\dodoi}[1]{doi:~\href{http://doi.org/#1}{\nolinkurl{#1}}}
\providecommand{\doeprint}[1]{\href{http://ascl.net/#1}{\nolinkurl{http://ascl.net/#1}}}
\providecommand{\doarXiv}[1]{\href{https://arxiv.org/abs/#1}{\nolinkurl{https://arxiv.org/abs/#1}}}

\bibitem[{{Alastuey} \& {Jancovici}(1978)}]{alastuey1978}
{Alastuey}, A., \& {Jancovici}, B. 1978, \apj, 226, 1034,
  \dodoi{10.1086/156681}

\bibitem[{{AMReX-Astro~initial\_models~Team}
  {et~al.}(2024){AMReX-Astro~initial\_models~Team}, Boyd, Smith~Clark, Willcox,
  \& Zingale}]{initial_models2024}
{AMReX-Astro~initial\_models~Team}, Boyd, B., Smith~Clark, A., Willcox, D., \&
  Zingale, M. 2024, AMReX-Astro/initial\_models: Release 24.03, 24.03,  Zenodo,
  \dodoi{10.5281/zenodo.10827428}

\bibitem[{{AMReX-Astro~Microphysics~Development~Team}
  {et~al.}(2024){AMReX-Astro~Microphysics~Development~Team}, Bishop, Fields,
  Chen, Harpole, Jacobs, Johnson, Katz, Li, Malone, Timmes, Willcox, \&
  Zingale}]{microphysics2024}
{AMReX-Astro~Microphysics~Development~Team}, Bishop, A., Fields, C.~E.,
  {et~al.} 2024, AMReX-Astro/Microphysics: Release 24.03, 24.03,  Zenodo,
  \dodoi{10.5281/zenodo.10732065}

\bibitem[{{Andrassy} {et~al.}(2022){Andrassy}, {Higl}, {Mao}, {Moc{\'a}k},
  {Vlaykov}, {Arnett}, {Baraffe}, {Campbell}, {Constantino}, {Edelmann},
  {Goffrey}, {Guillet}, {Herwig}, {Hirschi}, {Horst}, {Leidi}, {Meakin},
  {Pratt}, {Rizzuti}, {R{\"o}pke}, \& {Woodward}}]{andrassy2022}
{Andrassy}, R., {Higl}, J., {Mao}, H., {et~al.} 2022, \aap, 659, A193,
  \dodoi{10.1051/0004-6361/202142557}

\bibitem[{{Chamulak} {et~al.}(2008){Chamulak}, {Brown}, {Timmes}, \&
  {Dupczak}}]{chamulak2008}
{Chamulak}, D.~A., {Brown}, E.~F., {Timmes}, F.~X., \& {Dupczak}, K. 2008,
  \apj, 677, 160, \dodoi{10.1086/528944}

\bibitem[{{Cyburt} {et~al.}(2010){Cyburt}, {Amthor}, {Ferguson}, {Meisel},
  {Smith}, {Warren}, {Heger}, {Hoffman}, {Rauscher}, {Sakharuk}, {Schatz},
  {Thielemann}, \& {Wiescher}}]{cyburt2010}
{Cyburt}, R.~H., {Amthor}, A.~M., {Ferguson}, R., {et~al.} 2010, \apjs, 189,
  240, \dodoi{10.1088/0067-0049/189/1/240}

\bibitem[{{Denissenkov} {et~al.}(2013){Denissenkov}, {Herwig}, {Truran}, \&
  {Paxton}}]{denissenkov2013}
{Denissenkov}, P.~A., {Herwig}, F., {Truran}, J.~W., \& {Paxton}, B. 2013,
  \apj, 772, 37, \dodoi{10.1088/0004-637X/772/1/37}

\bibitem[{{Denissenkov} {et~al.}(2015){Denissenkov}, {Truran}, {Herwig},
  {Jones}, {Paxton}, {Nomoto}, {Suzuki}, \& {Toki}}]{denissenkov2015}
{Denissenkov}, P.~A., {Truran}, J.~W., {Herwig}, F., {et~al.} 2015, \mnras,
  447, 2696, \dodoi{10.1093/mnras/stu2589}

\bibitem[{{Fan} {et~al.}(2019{\natexlab{a}}){Fan}, {Nonaka}, {Almgren},
  {Willcox}, {Harpole}, \& {Zingale}}]{maestroex_joss}
{Fan}, D., {Nonaka}, A., {Almgren}, A., {et~al.} 2019{\natexlab{a}}, The
  Journal of Open Source Software, 4, 1757, \dodoi{10.21105/joss.01757}

\bibitem[{{Fan} {et~al.}(2019{\natexlab{b}}){Fan}, {Nonaka}, {Almgren},
  {Harpole}, \& {Zingale}}]{fan2019}
{Fan}, D., {Nonaka}, A., {Almgren}, A.~S., {Harpole}, A., \& {Zingale}, M.
  2019{\natexlab{b}}, \apj, 887, 212, \dodoi{10.3847/1538-4357/ab4f75}

\bibitem[{{Gilet} {et~al.}(2013){Gilet}, {Almgren}, {Bell}, {Nonaka},
  {Woosley}, \& {Zingale}}]{gilet2013}
{Gilet}, C., {Almgren}, A.~S., {Bell}, J.~B., {et~al.} 2013, \apj, 773, 137,
  \dodoi{10.1088/0004-637X/773/2/137}

\bibitem[{{Graboske} {et~al.}(1973){Graboske}, {Dewitt}, {Grossman}, \&
  {Cooper}}]{graboske1973}
{Graboske}, H.~C., {Dewitt}, H.~E., {Grossman}, A.~S., \& {Cooper}, M.~S. 1973,
  \apj, 181, 457, \dodoi{10.1086/152062}

\bibitem[{{Hansen} {et~al.}(2004){Hansen}, {Kawaler}, \& {Trimble}}]{HKT2004}
{Hansen}, C.~J., {Kawaler}, S.~D., \& {Trimble}, V. 2004, {Stellar interiors :
  physical principles, structure, and evolution} (Springer New York, NY),
  \dodoi{10.1007/978-1-4419-9110-2}

\bibitem[{Harris {et~al.}(2020)Harris, Millman, van~der Walt, Gommers,
  Virtanen, Cournapeau, Wieser, Taylor, Berg, Smith, Kern, Picus, Hoyer, van
  Kerkwijk, Brett, Haldane, del R{\'{i}}o, Wiebe, Peterson,
  G{\'{e}}rard-Marchant, Sheppard, Reddy, Weckesser, Abbasi, Gohlke, \&
  Oliphant}]{numpy2020}
Harris, C.~R., Millman, K.~J., van~der Walt, S.~J., {et~al.} 2020, Nature, 585,
  357, \dodoi{10.1038/s41586-020-2649-2}

\bibitem[{{Herwig} {et~al.}(2023){Herwig}, {Woodward}, {Mao}, {Thompson},
  {Denissenkov}, {Lau}, {Blouin}, {Andrassy}, \& {Paul}}]{herwig2023}
{Herwig}, F., {Woodward}, P.~R., {Mao}, H., {et~al.} 2023, \mnras, 525, 1601,
  \dodoi{10.1093/mnras/stad2157}

\bibitem[{{H{\"o}flich}(2006)}]{hoeflich2006}
{H{\"o}flich}, P. 2006, \nphysa, 777, 579,
  \dodoi{10.1016/j.nuclphysa.2004.12.038}

\bibitem[{{Hoyle} \& {Fowler}(1960)}]{hoyle1960}
{Hoyle}, F., \& {Fowler}, W.~A. 1960, \apj, 132, 565, \dodoi{10.1086/146963}

\bibitem[{Hunter(2007)}]{Hunter:2007}
Hunter, J.~D. 2007, Computing in Science and Engg., 9, 90,
  \dodoi{10.1109/MCSE.2007.55}

\bibitem[{{Itoh} {et~al.}(1996){Itoh}, {Nishikawa}, \& {Kohyama}}]{itoh1996}
{Itoh}, N., {Nishikawa}, A., \& {Kohyama}, Y. 1996, \apj, 470, 1015,
  \dodoi{10.1086/177926}

\bibitem[{{Itoh} {et~al.}(1979){Itoh}, {Totsuji}, {Ichimaru}, \&
  {Dewitt}}]{itoh1979}
{Itoh}, N., {Totsuji}, H., {Ichimaru}, S., \& {Dewitt}, H.~E. 1979, \apj, 234,
  1079, \dodoi{10.1086/157590}

\bibitem[{{Klein} \& {Pauluis}(2012)}]{klein2012}
{Klein}, R., \& {Pauluis}, O. 2012, Journal of the Atmospheric Sciences, 69,
  961, \dodoi{10.1175/JAS-D-11-0110.1}

\bibitem[{{Kritsuk} {et~al.}(2007){Kritsuk}, {Norman}, {Padoan}, \&
  {Wagner}}]{kritsuk2007}
{Kritsuk}, A.~G., {Norman}, M.~L., {Padoan}, P., \& {Wagner}, R. 2007, \apj,
  665, 416, \dodoi{10.1086/519443}

\bibitem[{{Langanke} \& {Wiescher}(2001)}]{langanke2001}
{Langanke}, K., \& {Wiescher}, M. 2001, Reports on Progress in Physics, 64,
  1657, \dodoi{10.1088/0034-4885/64/12/202}

\bibitem[{{Lesaffre} {et~al.}(2005){Lesaffre}, {Podsiadlowski}, \&
  {Tout}}]{lesaffre2005}
{Lesaffre}, P., {Podsiadlowski}, P., \& {Tout}, C.~A. 2005, \mnras, 356, 131,
  \dodoi{10.1111/j.1365-2966.2004.08428.x}

\bibitem[{{Liu} {et~al.}(2023){Liu}, {R{\"o}pke}, \& {Han}}]{liu2023}
{Liu}, Z.-W., {R{\"o}pke}, F.~K., \& {Han}, Z. 2023, Research in Astronomy and
  Astrophysics, 23, 082001, \dodoi{10.1088/1674-4527/acd89e}

\bibitem[{{Maoz} {et~al.}(2014){Maoz}, {Mannucci}, \& {Nelemans}}]{maoz2014}
{Maoz}, D., {Mannucci}, F., \& {Nelemans}, G. 2014, \araa, 52, 107,
  \dodoi{10.1146/annurev-astro-082812-141031}

\bibitem[{{Mart{\'{i}}nez-Rodr{\'{i}}guez}
  {et~al.}(2016){Mart{\'{i}}nez-Rodr{\'{i}}guez}, {Piro}, {Schwab}, \&
  {Badenes}}]{martinez-rodriguez2016}
{Mart{\'{i}}nez-Rodr{\'{i}}guez}, H., {Piro}, A.~L., {Schwab}, J., \&
  {Badenes}, C. 2016, \apj, 825, 57, \dodoi{10.3847/0004-637X/825/1/57}

\bibitem[{{Meakin} \& {Arnett}(2007)}]{meakin2007}
{Meakin}, C.~A., \& {Arnett}, D. 2007, \apj, 667, 448, \dodoi{10.1086/520318}

\bibitem[{Meurer {et~al.}(2017)Meurer, Smith, Paprocki, \v{C}ert\'{i}k,
  Kirpichev, Rocklin, Kumar, Ivanov, Moore, Singh, Rathnayake, Vig, Granger,
  Muller, Bonazzi, Gupta, Vats, Johansson, Pedregosa, Curry, Terrel,
  Rou\v{c}ka, Saboo, Fernando, Kulal, Cimrman, \& Scopatz}]{sympy}
Meurer, A., Smith, C.~P., Paprocki, M., {et~al.} 2017, PeerJ Computer Science,
  3, e103, \dodoi{10.7717/peerj-cs.103}

\bibitem[{{Nomoto} {et~al.}(1984){Nomoto}, {Thielemann}, \&
  {Yokoi}}]{nomoto1984}
{Nomoto}, K., {Thielemann}, F.~K., \& {Yokoi}, K. 1984, \apj, 286, 644,
  \dodoi{10.1086/162639}

\bibitem[{{Nonaka} {et~al.}(2012){Nonaka}, {Aspden}, {Zingale}, {Almgren},
  {Bell}, \& {Woosley}}]{nonaka2012}
{Nonaka}, A., {Aspden}, A.~J., {Zingale}, M., {et~al.} 2012, \apj, 745, 73,
  \dodoi{10.1088/0004-637X/745/1/73}

\bibitem[{{Paczy{\'n}ski}(1972)}]{paczynski1972}
{Paczy{\'n}ski}, B. 1972, \aplett, 11, 53

\bibitem[{{Paxton} {et~al.}(2013){Paxton}, {Cantiello}, {Arras}, {Bildsten},
  {Brown}, {Dotter}, {Mankovich}, {Montgomery}, {Stello}, {Timmes}, \&
  {Townsend}}]{paxton2013}
{Paxton}, B., {Cantiello}, M., {Arras}, P., {et~al.} 2013, \apjs, 208, 4,
  \dodoi{10.1088/0067-0049/208/1/4}

\bibitem[{{Perlmutter} {et~al.}(1999){Perlmutter}, {Aldering}, {Goldhaber},
  {Knop}, {Nugent}, {Castro}, {Deustua}, {Fabbro}, {Goobar}, {Groom}, {Hook},
  {Kim}, {Kim}, {Lee}, {Nunes}, {Pain}, {Pennypacker}, {Quimby}, {Lidman},
  {Ellis}, {Irwin}, {McMahon}, {Ruiz-Lapuente}, {Walton}, {Schaefer}, {Boyle},
  {Filippenko}, {Matheson}, {Fruchter}, {Panagia}, {Newberg}, {Couch}, \& {The
  Supernova Cosmology Project}}]{perlmutter1999}
{Perlmutter}, S., {Aldering}, G., {Goldhaber}, G., {et~al.} 1999, \apj, 517,
  565

\bibitem[{{Phillips}(1993)}]{phillips1993}
{Phillips}, M.~M. 1993, \apjl, 413, L105

\bibitem[{{Piersanti} {et~al.}(2017){Piersanti}, {Bravo}, {Cristallo},
  {Dom{\'\i}nguez}, {Straniero}, {Tornamb{\'e}}, \&
  {Mart{\'\i}nez-Pinedo}}]{piersanti2017}
{Piersanti}, L., {Bravo}, E., {Cristallo}, S., {et~al.} 2017, \apjl, 836, L9,
  \dodoi{10.3847/2041-8213/aa5c7e}

\bibitem[{{Piersanti} {et~al.}(2022){Piersanti}, {Bravo}, {Straniero},
  {Cristallo}, \& {Dom{\'\i}nguez}}]{piersanti2022}
{Piersanti}, L., {Bravo}, E., {Straniero}, O., {Cristallo}, S., \&
  {Dom{\'\i}nguez}, I. 2022, \apj, 926, 103, \dodoi{10.3847/1538-4357/ac403b}

\bibitem[{{Piro} \& {Bildsten}(2008)}]{piro2008}
{Piro}, A.~L., \& {Bildsten}, L. 2008, \apj, 673, 1009, \dodoi{10.1086/524189}

\bibitem[{{Riess} {et~al.}(1998){Riess}, {Filippenko}, {Challis},
  {Clocchiatti}, {Diercks}, {Garnavich}, {Gilliland}, {Hogan}, {Jha},
  {Kirshner}, {Leibundgut}, {Phillips}, {Reiss}, {Schmidt}, {Schommer},
  {Smith}, {Spyromilio}, {Stubbs}, {Suntzeff}, \& {Tonry}}]{riess1998}
{Riess}, A.~G., {Filippenko}, A.~V., {Challis}, P., {et~al.} 1998, \aj, 116,
  1009

\bibitem[{{Schwab} {et~al.}(2017){Schwab}, {Mart{\'{i}}nez-Rodr{\'{i}}guez},
  {Piro}, \& {Badenes}}]{schwab2017a}
{Schwab}, J., {Mart{\'{i}}nez-Rodr{\'{i}}guez}, H., {Piro}, A.~L., \&
  {Badenes}, C. 2017, \apj, 851, 105, \dodoi{10.3847/1538-4357/aa9a3c}

\bibitem[{Smith {et~al.}(2023)Smith, Johnson, Chen, Eiden, Willcox, Boyd, Cao,
  DeGrendele, \& Zingale}]{smith2023}
Smith, A.~I., Johnson, E.~T., Chen, Z., {et~al.} 2023, The Astrophysical
  Journal, 947, 65, \dodoi{10.3847/1538-4357/acbaff}

\bibitem[{{Stein} \& {Wheeler}(2006)}]{stein-wheeler2006}
{Stein}, J., \& {Wheeler}, J.~C. 2006, \apj, 643, 1190, \dodoi{10.1086/503246}

\bibitem[{{Suzuki} {et~al.}(2016){Suzuki}, {Toki}, \& {Nomoto}}]{suzuki2016}
{Suzuki}, T., {Toki}, H., \& {Nomoto}, K. 2016, \apj, 817, 163,
  \dodoi{10.3847/0004-637X/817/2/163}

\bibitem[{{Timmes} \& {Swesty}(2000)}]{timmes2000}
{Timmes}, F.~X., \& {Swesty}, F.~D. 2000, \apjs, 126, 501,
  \dodoi{10.1086/313304}

\bibitem[{{Turk} {et~al.}(2011){Turk}, {Smith}, {Oishi}, {Skory}, {Skillman},
  {Abel}, \& {Norman}}]{turk2011}
{Turk}, M.~J., {Smith}, B.~D., {Oishi}, J.~S., {et~al.} 2011, The Astrophysical
  Journal Supplement Series, 192, 9, \dodoi{10.1088/0067-0049/192/1/9}

\bibitem[{{Vasil} {et~al.}(2013){Vasil}, {Lecoanet}, {Brown}, {Wood}, \&
  {Zweibel}}]{vasil2013}
{Vasil}, G.~M., {Lecoanet}, D., {Brown}, B.~P., {Wood}, T.~S., \& {Zweibel},
  E.~G. 2013, \apj, 773, 169, \dodoi{10.1088/0004-637X/773/2/169}

\bibitem[{{Willcox}(2018)}]{willcox2018}
{Willcox}, D.~E. 2018, PhD thesis, SUNY Stony Brook, New York

\bibitem[{{Willcox} \& {Zingale}(2018)}]{pynucastro}
{Willcox}, D.~E., \& {Zingale}, M. 2018, Journal of Open Source Software, 3,
  588, \dodoi{10.21105/joss.00588}

\bibitem[{{Woosley} {et~al.}(1986){Woosley}, {Taam}, \& {Weaver}}]{woosley1986}
{Woosley}, S.~E., {Taam}, R.~E., \& {Weaver}, T.~A. 1986, \apj, 301, 601,
  \dodoi{10.1086/163926}

\bibitem[{{Woosley} {et~al.}(2004){Woosley}, {Wunsch}, \&
  {Kuhlen}}]{woosley2004}
{Woosley}, S.~E., {Wunsch}, S., \& {Kuhlen}, M. 2004, \apj, 607, 921,
  \dodoi{10.1086/383530}

\bibitem[{Zhang {et~al.}(2019)Zhang, Almgren, Beckner, Bell, Blaschke, Chan,
  Day, Friesen, Gott, Graves, Katz, Myers, Nguyen, Nonaka, Rosso, Williams, \&
  Zingale}]{zhang2019}
Zhang, W., Almgren, A., Beckner, V., {et~al.} 2019, Journal of Open Source
  Software, 4, 1370, \dodoi{10.21105/joss.01370}

\bibitem[{{Zingale} {et~al.}(2009){Zingale}, {Almgren}, {Bell}, {Nonaka}, \&
  {Woosley}}]{zingale2009}
{Zingale}, M., {Almgren}, A.~S., {Bell}, J.~B., {Nonaka}, A., \& {Woosley},
  S.~E. 2009, \apj, 704, 196, \dodoi{10.1088/0004-637X/704/1/196}

\end{thebibliography}
\bibliographystyle{aasjournal}

\end{document}